\documentclass{aa}

\usepackage{graphicx}
\usepackage{txfonts}
\usepackage{xcolor}
\usepackage[normalem]{ulem}
\usepackage{hyperref}
\begin{document}


\title{
Viscous torque in turbulent magnetized AGN accretion disks and its effects on EMRI's gravitational waves}

   \author{ Fatemeh Hossein Nouri
          \and
          Agnieszka Janiuk
          }

   \institute{Center for Theoretical Physics, Polish Academy of Sciences, Al. Lotnikow 32/46, 02-668 Warsaw, Poland\\
              \email{f.h.noori@cft.edu.pl}
             }



\abstract
{The merger of supermassive black holes (SMBHs) produces mHz gravitational waves (GW), which are potentially detectable by future Laser Interferometer Space Antenna (LISA). Such binary systems are usually embedded in an accretion disk environment at the centre of the active galactic nucleus (AGN). Recent studies suggest the plasma environment imposes measurable imprints on the GW signal if the mass ratio of the binary is around q $ \sim10^{-4}-10^{-3}$. The effect of the gaseous environment on the GW signal is strongly dependent on the disk’s parameters, therefore it is believed that future low-frequency GW detections will provide us with precious information about the physics of AGN accretion disks. We investigate this effect by measuring the viscous torque via modelling the evolution of magnetized tori around the primary massive black hole. Using GRMHD HARM-COOL code, we perform 2D and 3D simulations of weakly-magnetized thin accretion disks, with a possible truncation and transition to advection-dominated accretion flow (ADAF). We study the angular momentum transport and turbulence generated by the magnetorotational instability (MRI). We quantify the disk’s effective alpha viscosity and its evolution over time. We apply our numerical results to quantify the relativistic viscous torque on a hypothetical low-mass secondary black hole via 1D analytical approach, and we estimate the GW phase shift due to the gas environment.}


\titlerunning{Disk's environmental effects on GW from EMRIs}

\maketitle


\section{\label{sec:intro}Introduction}

Binary supermassive black holes (SMBHs) are expected to be formed in galaxy mergers where their corresponding SMBHs pair up into binaries. Such binary systems are usually embedded in a gaseous environment at the center of active galactic nuclei (AGN). Stellar dynamical friction and torques from gas are expected to bring the binary to the sub-parsec scale~\citep{Milosavljevic:2003}. In about a sub-parsec binary separation, the system emits low-frequency gravitational waves ($\sim$mHz) through the inspiral and merger phases, which are possibly detectable by future observatories such as the Laser Interferometric Space Antenna (LISA)~\citep{LISA:2023}. So far, several candidate binary SMBHs have been identified in AGN and quasars using different observational methods, including NuSTAR and Chandra X-ray observations~\citep{Saade:2020,Saade:2023} and periodic features in AGN light curves~\citep{Graham:2015}. The current strongest candidates are OJ 287 and PG 1302-102, which both display periodicity in their lightcurves~\citep{Bogdanovic:2022}.

The accretion of gas by the binary may result in appreciable electromagnetic (EM) radiation. On the other hand, the orbital frequency of the binary can be affected by the torque imposed by the gas environment and may cause a phase shift in the GW signal. 
The EM emissions and the GW signal together may provide us with a useful probe of the gas in galaxy cores and a diagnostic of the physics of black hole accretion. In recent years, several analytical and numerical studies have been done to study this electromagnetic emission mostly for a binary system with equal masses in general relativistic hydrodynamics~\citep{Farris:2011,Shapiro:2013} and general relativistic magnetohydrodynamics (GR MHD)~\citep{Noble:2012,DAscoli:2018,Bowen:2018,Combi:2022,Avara:2023}.

Several groups have studied the interaction of the gaseous circumbinary disk with the binary system via the 2D and 3D nonrelativistic hydrodynamic simulations. They include the viscous gas described by the Navier-Stokes fluid equations, interacting with an equal mass circular binary, assuming the binary orbit to remain fixed ~\citep{Moody:2019,Tiede:2020,Mahesh:2023} or evolving with time~\citep{Franchini:2022,Franchini:2023}. Depending on the assumptions for the live binary orbit, grid resolution and circumbinary disk parameters, they found that the exchange of angular momentum between the disk and the binary may lead to binary expansion or shrinkage. 

Although the equal mass binary studies are more common in the literature, the GW phase shift due to the environmental effect is more significant for the SMBHs with extreme mass ratio inspirals (EMRIs) ($q \sim 10^{-4}-10^{-3}$) and more probable to be detected with the future laser interferometric observatories~\citep{Barausse:2014}. 
Based on semi-analytical studies by~\cite{Yunes:2011} and~\cite{Kocsis:2011} the accretion disk environment imposes measurable imprints on the GW signal in the EMRIs. 
These studies show that depending on the disk parameters, the perturbation of the GW phase is between 10 and 1000 radians per year, detectable by LISA. 
\cite{Barausse:2008} explored the effects of BH spin (including prograde and retrograde orbits), disk's mass and density on the orbital evolution of the EMRIs.
\cite{Derdzinski:2020} performed 2D nonrelativistic numerical simulations inspired by the work of~\cite{Duffell:2014} for measuring the torques on a Jupiter-like planet, embedded in a protoplanetary disk. They applied the same code to the AGN context and found that if the disk's surface density is high enough, the phase shift exceeds a few radians in a 5-year LISA observation. 
Most recently,~\cite{Garg:2022} have done a parameterized study on the gas torque measurements in an analytical approach for intermediate-mass black hole binaries embedded in $\alpha$-disks. They quantified the torque over a wide range of Shakura-Sunyaev disk's characteristics such as surface density and Mach number, as well as primary BH's mass and the binary's mass ratio.   
\cite{Cardoso:2020} and~\cite{Speri:2023} determined how well the environmental effects can be measured using gravitational wave observations from LISA, and a new fully relativistic formalism has been introduced by~\cite{Cardoso:2022} to study gravitational wave emission by EMRIs in non-vacuum curved spacetimes.

In this paper, we focus on the case of extreme mass ratio binaries. We simulate a geometrically thin, Keplerian disk orbiting in the plane of a spinning BH. The low-mass companion BH is explicitly not included in our numerical simulation, instead, we use an analytic estimation to measure the viscous torque on the secondary BH. The radiative cooling is neglected in our simulations, therefore our models are applicable to the radiatively inefficient accretion flows~\citep{Narayan&Yi:1994}, e.g. low-luminous AGN, such as NGC 5548~\citep{Bentz:2007,Crenshaw:2009}. 
Interestingly, in this source the observed profile variability of broad emission lines \citep{shapovalova} may suggest an inspiral effect, and formation of a hot spot
due to a collision of the disc with a passing star, 
or a binary black hole system \cite{bon}.

The previous works addressed to the environmental effects of the disk onto GW signal assumed the artificial $\alpha$-prescription as the mechanism for the angular momentum transport~\citep{Shakura:1976}. In this approach, the $\alpha$ viscosity is assumed a constant value (typically, $\sim$ 0.01-0.1) for the entire disk and for the entire time of the inspiral phase. 
However, in a more realistic approach, one needs to include the magnetic field evolution to provide the physical mechanism for the angular momentum transport caused by the magnetorotational instability (MRI)~\citep{BalbusHaw1991}. To include this realistic assumption, we perform a full general relativistic MHD simulation. We seed the initial magnetic field to have a weakly magnetized plasma and evolved the disk in time. We quantify the effective value of $\alpha$ directly from the Reynolds and Maxwell contribution to the stress-energy tensor, as computed for different parts of the disk and over time.
Most our simulations are axisymmetric, but to check the importance of non-axisymmetric effects, we also perform a 3D simulation. As the measured torque varies over time and radius, our calculations impose new constraints on the GW phase shift estimation derived from the $\alpha$ disk approach. We evolve the system with GR MHD equations, therefore our study includes the curved spacetime and black hole spin effects. 
In addition to the spin, we study the effects of other physical parameters such as magnetic field strength, and its configuration. 

The paper is organized as follows.
In Section~\ref{sec:setups} we describe the initial configuration of our simulation and the unit conversions. The numerical results are presented in Sec. \ref{sec:results} with a detailed discussion on MRI analysis and $\alpha$-parameter computation. Sec. \ref{sec:torque} is devoted to our estimations of the viscous torque and its fluctuations. In Sec. \ref{sec:discussion} we discuss the importance of our results on the future GW detection by LISA with a rough estimation of possible dephasing due to the gaseous environment, and we give a brief comparison of our $\alpha$ values with the previous studies. Finally, the summary and conclusions are given in Sec.~\ref{sec:conclusion}. 

\section{\label{sec:setups}Methods and Setups}
\subsection{Numerical methods and initial configuration}

We use our version of the GR MHD code HARM described in~\cite{Janiuk2017, Sap2019}, which uses numerical algorithms developed initially by~\cite{Gammie2003} and~\cite{Noble2006}. HARM uses a conservative shock-capturing scheme, with fluxes calculated using classical Harten-Lax-van Leer method. The constrained transport maintains divergence free magnetic field.  
The background spacetime is fixed by the Kerr metric of the black hole with constant mass and angular momentum. The hydro equations are evolved in the modified spherical Kerr-Schild coordinates that are non-singular on the horizon. The following radial and angular maps from Kerr-Schild (KS) coordinates to modified Kerr-Schild (MKS) coordinates are used, which increase the resolution close to the black hole and the equatorial plane, respectively, to resolve the thin disk accurately: $r_{KS} = \exp({r_{MKS}})$, $\theta_{KS} = \pi \theta_{MKS} + \frac{(1-h)}{2}\sin(2\pi \theta_{MKS})$. The coordinate parameter $h$ is set to $0.3$ for all models in this paper.

The gas pressure is calculated using a polytropic equation of state $P=\kappa \rho^{\Gamma}$, with $\kappa=0.1$, and $\Gamma=4/3$.
The initial density configuration is based on~\cite{Dihingia:2021} prescription for thin disks. The distribution of density on the equator is defined as:

\begin{equation}
    \rho_e = \left( \frac{\Theta_0}{\kappa} \right)^{\frac{1}{(\Gamma -1)}} \left( \frac{f(x)}{x^2} \right)^{\frac{1}{4(\Gamma - 1)}}.
\end{equation}

The disk is truncated at the innermost stable circular orbit radius, $r_{\textit{ISCO}}$, and the density on the entire grid is derived from: 

\begin{equation}
    \rho (r,\theta) = \rho_e~\textit{exp}\left( -\frac{\alpha_{disk}^2 z^2}{\mathcal{H}^2} \right);~z = r~cos(\theta).
\end{equation}

The $\Theta_0$ parameter is the dimensionless temperature set to $\Theta_0=0.001$, the parameter $\alpha_{disk} = 2$ is chosen to keep the disk thin ($H/R \approx 0.05$), and $x=\sqrt{r}$. We refer the readers to eqs.(4-13) from~\cite{Dihingia:2021} for definitions of $f(x)$ and $\mathcal{H}$.

The initial poloidal magnetic field configuration is based on~\cite{Zanni:2007} with the nonzero azimuthal component of the magnetic vector potential defined as:

\begin{equation}
    A_{\phi} = r^{3/4} \frac{m^{5/4}}{(m^2 + cos^2 \theta)^{5/8}}. 
\end{equation}

\begin{figure}
\centering
\includegraphics[width=0.49\textwidth]{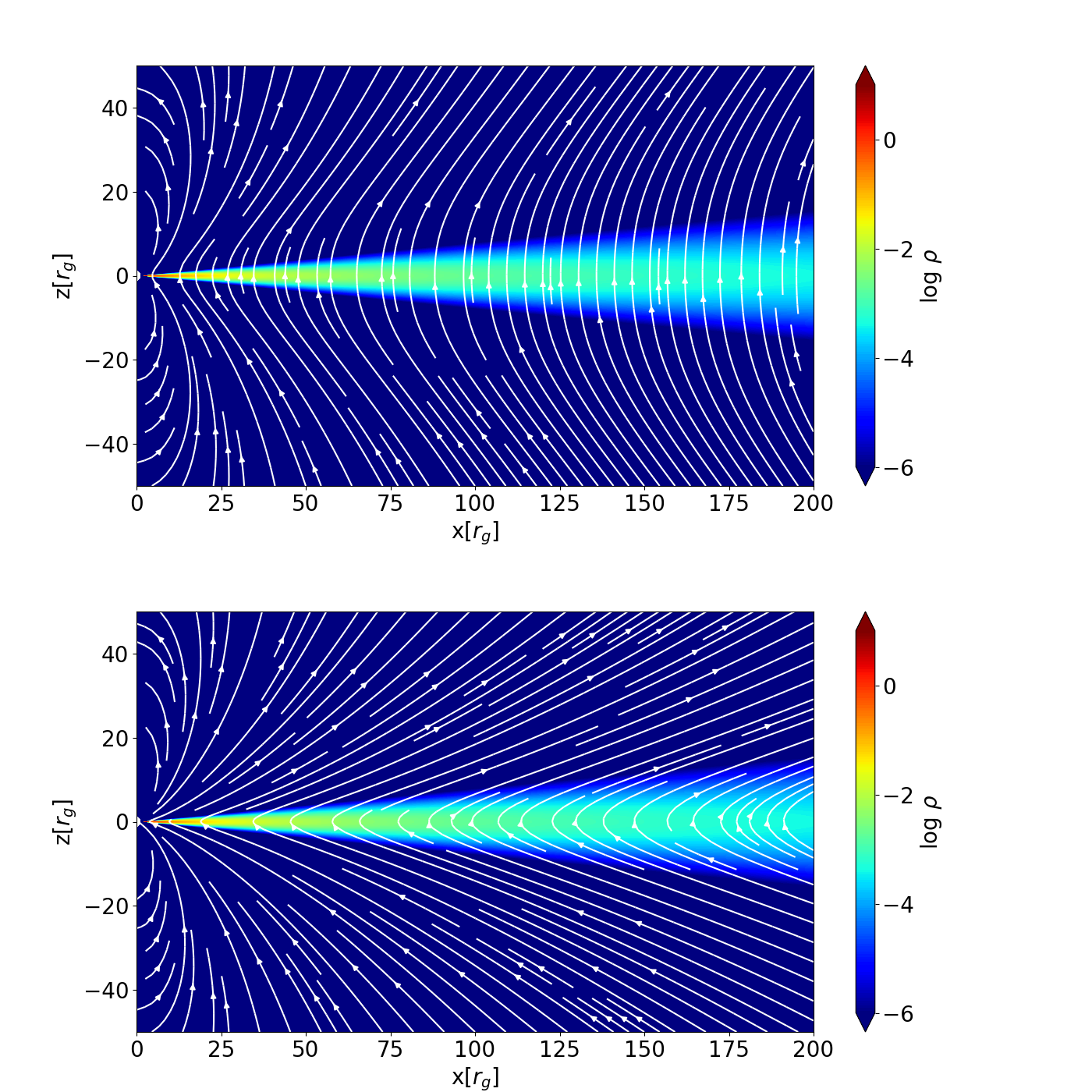}
\caption{The 2D profile of density and magnetic field lines at the initial time, for $\beta$50-m0.5-a0.94 (top) and $\beta$50-m0.1-a0.94 cases (bottom).}
\label{fig:initial-setup}
\end{figure}

Here $m$ is a constant and defines the inclination angle of the initial magnetic field. For this study, we choose $m=0.1$ (low inclination angle) and $m=0.5$ (high inclination angle) for different cases as shown in Fig.~\ref{fig:initial-setup}.

\begin{figure}
\centering
\includegraphics[width=0.42\textwidth]{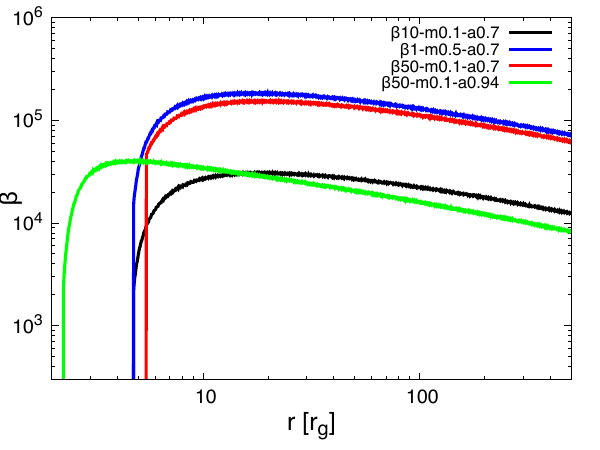}
\caption{The radial profile of $\beta$ parameter on the equator at the initial time for all the cases.}
\label{fig:beta-t0}
\end{figure}

We normalize the initial field strength with a given value of the ratio of the maximum gas pressure to the maximum magnetic pressure, $\beta=P_{gas}^{max}/P_{mag}^{max}$. The radial profile of $P_g/P_B$ ratio on the equator at the initial time is shown in Fig.~\ref{fig:beta-t0}.

We summarize the initial parameters of our simulations in Table~\ref{tab:test-cases}. The main simulations are performed in 2D with axisymmetric assumption and the grid resolution is 1056*528 points in radial and polar directions, respectively. We present a couple of test cases to study the higher resolution and the 3D model in Sec.\ref{sec:3D}.
The outer boundary is located at $r=1000$\,\textit{$r_g$}.
All cases are evolved for about $t \sim 60000$ \textit{$t_g$}.
The geometric units $r_g$ and $t_g$ and their relations with physical units are explained in Sec.~\ref{sec:phys_units}.

\begin{table}[]
    \centering
    \begin{tabular}{ l l l l l }
    \hline 
         case & $\beta$ & m & BH spin &  $\beta_{\textit{max, eq}}$ \\
         \hline 
         $\beta$1-m0.5-a0.7 & 1 & 0.5 & 0.7 & 185244  \\
         $\beta$10-m0.1-a0.7 & 10 & 0.1 & 0.7 & 31246 \\
         $\beta$50-m0.1-a0.7 & 50 & 0.1 & 0.7 & 156000 \\
         $\beta$50-m0.1-a0.94 & 50 & 0.1 & 0.94 & 39313 \\
    \hline      
    \end{tabular}
    \caption{The initial setup parameters for all the simulations with grid resolution: 1056*528.}
    \label{tab:test-cases}
\end{table}

\subsection{\label{sec:phys_units}Physical scales and physical units}

In HARM, we use geometric units: $G=c=M=1$. To convert quantities to physical units we follow~\cite{Janiuk-2019}, where the spacial and time units are scaled with the mass of the primary BH as follows:

\begin{equation}
\begin{aligned}
   & L_{unit} = \frac{GM}{c^2} = 1.48 \times 10^5 \frac{M}{M_{\odot}}cm, \\
   & T_{unit} = \frac{r_g}{c} = 4.9 \times 10^{-6} \frac{M}{M_{\odot}}s.
   \label{eq:unit}
\end{aligned}
\end{equation}

The density scale is related to the length unit by $\rho_{unit} = M_{scale}/L_{unit}^3$, and the mass scale is adopted to $M_{scale} = 1 \times 10^{-5} M_{\odot}$ for $\beta$50-m0.1-a0.94 case, and $M_{scale} = 2 \times 10^{-6} M_{\odot}$ for the rest of the models. The density scales are chosen to create disks with surface density $\Sigma \sim 10^3$ g cm$^{-2}$ around the radius of $r \sim 100 ~r_g$ as suggested by~\cite{Derdzinski:2020}.
These scaling factors give the measured accretion rate $\sim 0.01-1 ~\Dot{M}_{Edd}$ for our simulations (see Sec.~\ref{sec:results} for detailed discussion on the accretion rates).

If we assume the primary black hole has the mass of $M = 10^6 M_{\odot}$, the outer boundary is located at $r = 1000\,r_{g} \sim 1.5 \times 10^{12}$cm, and the evolution time is about $t \sim 60000\,t_g\sim 3.5$ days based on Eq.(\ref{eq:unit}).
According to these scales, we can claim that our model covers only the inner part of the AGN disks (which extends to $10^{14}-10^{16}$cm according to~\cite{frank_king_raine_2002}), and the evolution time covers only a fraction of the LISA observational time for SMBH inspiral ($\sim$ several years; ~\cite{Derdzinski:2020}). 
This evolution time is insufficient to make the entire disk turbulent for the selected resolution. Therefore, we consider only the region inside $r<200 ~r_g$ for our torque measurements (assuming the hypothetical secondary BH is orbiting over this range of binary separations). The fluid completes more than 20 orbits at this radius based on the Keplerian orbital frequency. 

According to observations, the Seyfert 2 galaxy GSN 069 at a redshift of $z = 0.018$ with nine-hour X-ray quasi-periodic eruptions is a candidate for extreme mass ratio binary black holes. The primary BH is a low mass SMBH of a few times $10^5 M_{\odot}$ with a relatively high Eddington ratio of about 0.5~\citep{Miniutti:2019}. The variability observed in GSN 069 may be explained by the interaction between an existing accretion disk and an orbiting secondary body according to~\cite{Linial-Metzger:2023,Franchini:2023QPE,Tagawa-Haiman:2023,Miniutti:2023}. However, another possible explanation has been suggested based on self-gravitational-lensing in SMBHs by~\cite{ingram21} for low-mass AGNs such as GSN 069 and RX J1301.9+2747. Our moderately high Eddington ratio models can resemble this type of object, therefore, we scale our results for such low-mass primary SMBH ($M_p \sim 10^5-10^6 M_{\odot}$) with an intermediate-mass companion BH.

\section{\label{sec:results}Numerical results}
\subsection{\label{sec:MRI}Disk evolution and MRI analysis}

\begin{figure*}
    \centering
    \includegraphics[width=0.95\textwidth]{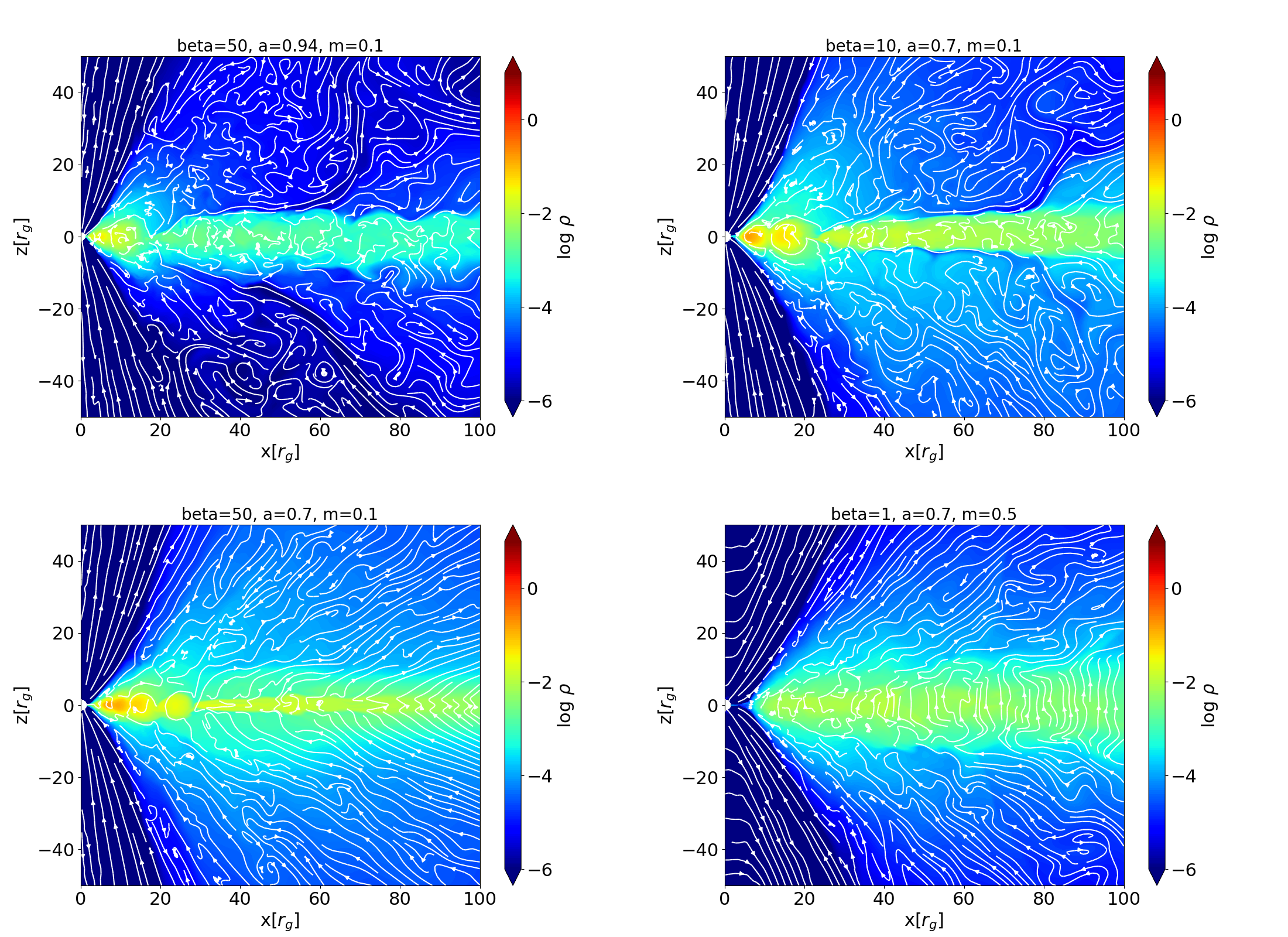}
    \caption{The 2D profiles of density and field lines at the final time for different models. The dense and strongly magnetized inner torus is visible in cases $\beta$50-m0.1-a0.94 and $\beta$10-m0.1-a0.7 separated from the turbulent thin disk. The $\beta$50-m0.1-a0.7 has multiple plasmoid structures at the inner part. The model $\beta$1-m0.5-a0.7 turns into the MAD state at the early time of the evolution, with the magnetic field preserved in a vertical configuration for most parts of the disk.}
    \label{fig:2D-rho-final}
\end{figure*}

We evolved our models for about $t \approx 60000$~$t_g$. This time is long enough to let the MRI be active to form the turbulent structure at the inner part of the disk ($r<150~r_g$), and short enough to avoid the magnetic field dissipation due to the anti-dynamo effect in a 2D simulation (the 3D simulation of $\beta$50-m0.1-a0.94 case supports this claim; see Sec.~\ref{sec:3D} for more details).
In Fig.~\ref{fig:2D-rho-final} we show the final configuration of the density and magnetic field lines at the end of the simulations for different models.

At the earlier evolution time, for cases $\beta$10-m0.1-a0.7, $\beta$50-m0.1-a0.7, and $\beta$50-m0.1-a0.94, the magnetic field is amplified exponentially due to the MRI and magnetic winding, which creates turbulent magnetized fluid with quite high accretion rates. 
As a result, the disk expands vertically, launches the outflows and becomes slightly thicker geometrically compared to the initial configuration. However, in these cases, the thin structure of the disk is preserved during evolution, creating a stream that flows radially toward the BH on the equator. This observation suggests that the MRI channel solution is developed in our simulations (channel solutions represent a specific form of poloidal MRI, which is characterised by prominent radial extended features~\citep{BalbusHaw1991,Stone-Norman:1994,Dibi:2012}).

\begin{figure*}
    \centering
    \includegraphics[width=0.97\textwidth]{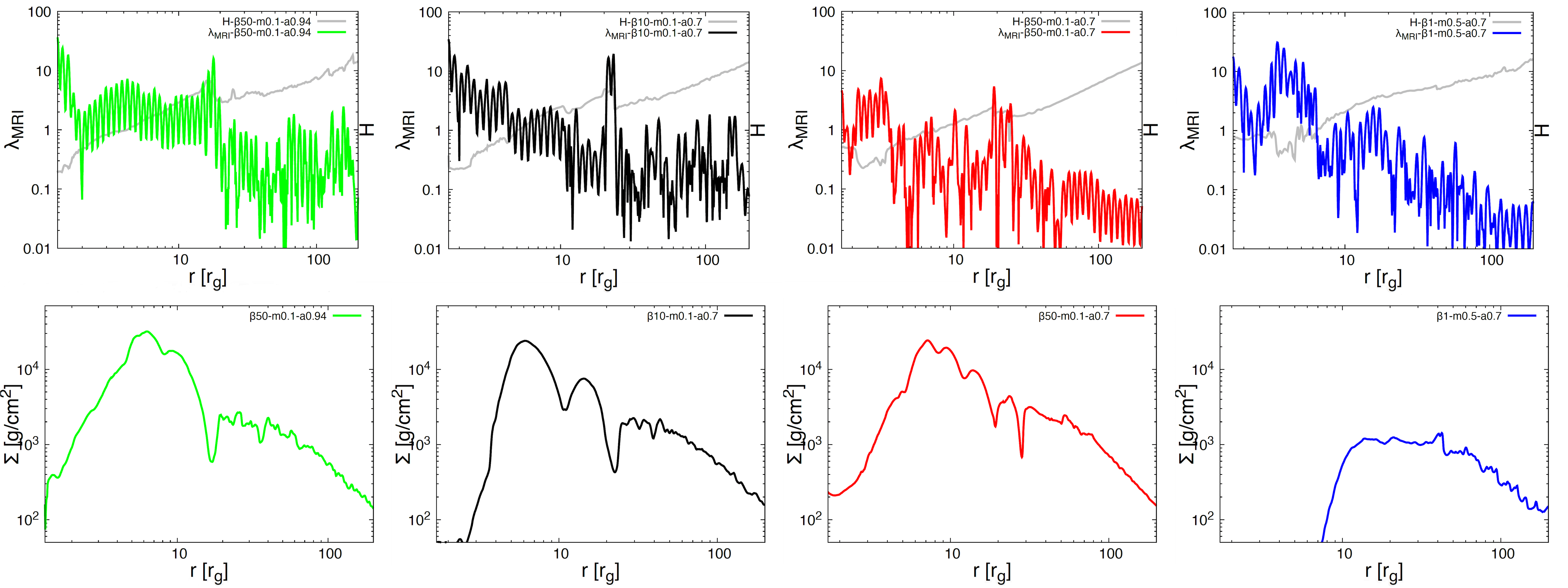}
    \caption{The comparison of the MRI fastest growing mode's wavelength with the disk scale height at the final time (top), and the surface density at the same time for all the models (bottom). The high-density plasmoid structures are formed at the inner radii, where the MRI is suppressed. The surface density is scaled for the central BH mass $10^6 M_{\odot}$.}
    \label{fig:MRI-sigma}
\end{figure*}

At later times, we observe that the disks are divided into two distinguishable parts (for these three cases). The inner region is denser, geometrically thicker and more magnetized, while the outer part is geometrically thinner, less dense and less magnetized. The formation of the inner 'mini-torus' has been observed before in magnetized thick disks in 2D simulations~\citep{DeVilliers:2003}. In order to investigate the MRI effect and the formation of the inner torus, we compare the radial profiles of the MRI fastest growing mode wavelength, $\lambda_{MRI}$, to the scale height of the disk in Fig.~\ref{fig:MRI-sigma}. The instability is suppressed when $\lambda_{MRI}$ exceeds the scale height~\citep{McKinney:2012,White:2019}, which happens for $r < 20 ~r_g$ in $\beta$50-m0.1-a0.94 case, where the mini-torus is formed. On the other hand, the visualized data show that the magnetic field lines loop inside the inner torus and form a magnetic barrier at its boundaries, and therefore, create a plasmoid structure. A detailed study on the formation of plasmoids due to magnetic reconnection and their observational effects should be done with high-resolution 3D simulations~\citep{Ripperda:2022}. With our current 2D moderate-resolution simulation we can explain the existence of the inner torus as a result of two physical processes intensifying each other: (i) the effective MRI at the outer radii makes the fluid lose its angular momentum and be dragged inwards, while the less effective MRI at the inner radii makes the hot magnetized fluid slow down and pile up over time, and (ii) at the same time the magnetic field is amplified and creates loops in the inner region causing the plasma trapped and disconnected from the rest of the disk. Therefore, the inner torus is formed and becomes stable till the end of the simulation. For $\beta$10-m0.1-a0.7 and $\beta$50-m0.1-a0.7 cases, multiple loop structures are visible. Fig.~\ref{fig:MRI-sigma} shows the comparison of the MRI fastest growing mode's wavelength $\lambda_{MRI}$, with the disk scale height $H$ at the final time. 
The scale height is defined as $H=c_s/\Omega_{rot}$, and the MRI fastest growing wavelength is $\lambda_{MRI} \approx 2\pi |v_{\theta,A}|/|\Omega_{rot}|$, where $c_s$ is the adiabatic sound speed, $\Omega_{rot}$ is the fluid rotational frequency, and $v_{\theta,A}$ is the $\theta$-directed Alfven velocity (cf. ~\cite{McKinney:2012}). 
We observe that for all models, the MRI is suppressed locally at the smaller radii, where the $\lambda_{MRI} > H$. The bottom panel of the same figure shows the surface density profile over the same radial region at the same time. The dense inner torus is distinguishable for $\beta$10-m0.1-a0.7, $\beta$50-m0.1-a0.7, and $\beta$50-m0.1-a0.94 models.

\begin{figure}
\centering
\begin{minipage}{0.42\textwidth}
\includegraphics[width=\textwidth]{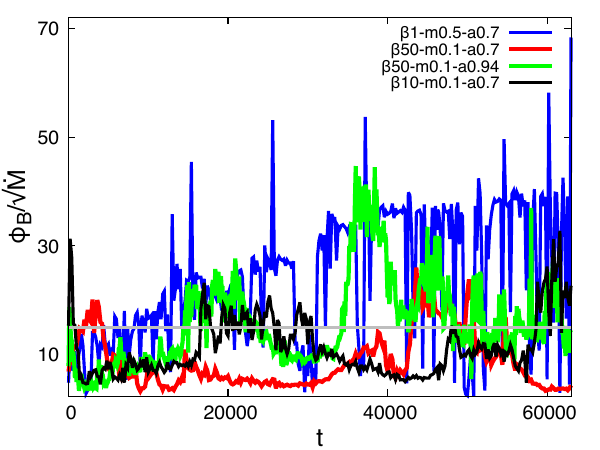}
\caption{The ratio of the magnetic flux to the square root of the mass flux computed at the BH horizon for all the cases. The disks turn to MAD state when this ratio goes above 15. The horizontal line shows the MAD criterion, i.e. ratio equals to 15.}
\label{fig:flux-ratio}
\end{minipage}
\quad
\begin{minipage}{0.42\textwidth}
\includegraphics[width=\textwidth]{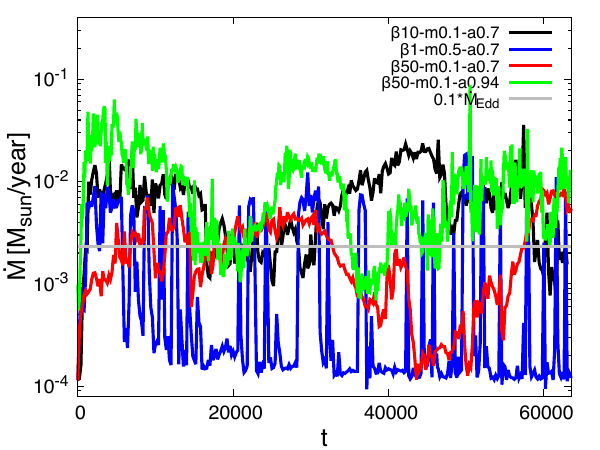}
\caption{The accretion rate for different cases compared with the Eddington accretion limit. The results are scaled for central BH with the mass of $10^6 M_{\odot}$.}
\label{fig:mdot}
\end{minipage}
\end{figure}

Completing our MRI analysis, we investigate the possibility of transition to a magnetically arrested disk (MAD), where the disk becomes magnetically dominated and the MRI is suppressed. The MAD status is considered a probable scenario for the disks at the center of a galaxy. The 3D numerical simulations done by~\cite{Liska:2020} suggested that for a long enough evolution, the disk eventually turns into the MAD state, and in this state, the final disk's characteristics are not sensitive to the exact initial conditions it started with. The recent observations by the Event Horizon Telescope confirm that the MAD state is more favourable for observed sources such as M87~\citep{EHT2021}. To investigate this, we measure the ratio of the magnetic flux to the square root of the mass flux at the BH horizon $\Phi_B/\sqrt{\dot{M}}$ for different cases. Based on the literature, the MAD state happens when this ratio is high enough, $ \sim 15$ ~\citep{Tchekhovskoy:2011}. Fig.~\ref{fig:flux-ratio} shows that $\beta$50-m0.1-a0.94 case, for instance, becomes magnetically arrested for a part of the evolution. At this period of time, the MRI does not act as an effective process for the angular momentum transport, and the accretion rate drops significantly as illustrated in Fig.~\ref{fig:mdot}. 
However, the accretion rate is maintained at around $\sim 10^{-2} M_{\odot}$/year for $\beta$10-m0.1-a0.7 and $\beta$50-m0.1-a0.94 cases at the end of the simulations, where the MAD condition is marginally satisfied or below the threshold, i.e. $\Phi_B/\sqrt{\dot{M}} \leq 15$.

At this point, we would like to highlight our $\beta$1-m0.5-a0.7 case, which started with a higher inclination angle for the magnetic field initial configuration. The changes are quite dramatic for this case, and it turns to MAD state at an earlier time and in an episodic way (see Figs.~\ref{fig:flux-ratio} and~\ref{fig:mdot}). Such episodic accretion rates are commonly observed in MADs~\citep{Igumenshchev:2008,Dihingia:2021}. More specifically, at the inner part, the magnetic winding causes high magnetic pressure and creates a magnetic barrier which reduces the accretion rate to below $0.01 \dot{M}_{Edd}$, while the other cases have accretion rate higher than $0.1 \dot{M}_{Edd}$ for a long period of evolution. At the further radii, as illustrated in Fig.~\ref{fig:MRI-sigma}, in a tiny fraction of the disk we have a turbulent structure with the $\lambda_{MRI}$ standing below the disk scale height and yet high enough to be resolved with current resolution. However, the vertical configuration of the magnetic field is preserved for the most part of the disk ($r>50\,r_g$) till the end of the simulation, and makes the disk expand vertically and become less dense compared to the other cases (see Fig.~\ref{fig:2D-rho-final}). To investigate this further, we performed a test with a similar setup but a weaker initial magnetic field, i.e. $\beta=50$. (The result of this test is not presented in our figures). 
The result shows that even with a weaker initial magnetic field, the MRI can not be triggered at larger radii and the GR MHD evolution keeps the magnetic field in vertical geometry in most parts of the disk. Therefore, the vertical configuration most likely results in either weak MRI or MAD state. Similar works in the literature confirm that a modest change in the initial field configuration may signify MAD structure~\citep{Narayan:2003,McKinney:2012,Dihingia:2021}.

Overall, the results presented in Figs.~\ref{fig:MRI-sigma} and ~\ref{fig:flux-ratio} determine that the MRI and its effects are not only suppressed locally at the inner part of the disk but are also suppressed globally during the evolution when the disk turns into the MAD state. 

\subsection{\label{sec:alpha}$\alpha$ measurement}

The MRI analysis presented in Sec.\ref{sec:MRI} shows that this instability does not remain active everywhere in the disk and for the entire time of the evolution. Therefore the viscous effects driven by MRI vary with radius and time as well. In this section, we compute the equivalent $\alpha$ viscosity caused by the MRI in the turbulent fluid and compare it with the constant values used in the literature. 

Following the prescription given by~\cite{McKinney:2012} for turbulent relativistic fluid, we calculate equivalent $\alpha$ viscosity by considering the dominant Reynolds and Maxwell terms in the stress-energy tensor as:

\begin{equation}
\begin{aligned}
    \alpha =  \alpha_{R} + \alpha_{M}, \\
    \alpha_{R} \approx \frac{\rho_{0} \delta u_r \delta u_{\phi}   \sqrt{g^{\phi \phi}}}{P_{tot}}, \\ 
    \alpha_{M} \approx - \frac{b_r b_{\phi} \sqrt{g^{\phi \phi}}}{P_{tot}}.
\end{aligned}
\end{equation}

In these equations $P_{tot} = P_{gas}+P_{mag}$ is the total pressure, and $b^{\mu}$ is the magnetic field 4-vector (see Eq.(8) from~\cite{Gammie2003} for the definition). The 4-velocity fluctuations are defined by:

\begin{equation}
\begin{aligned}
    \delta u_r(r,\theta,t) = u_r(r,\theta,t) - \overline{u_r(r,t)}, \\   
    \delta u_{\phi}(r,\theta,t) = u_{\phi}(r,\theta,t) - \overline{u_{\phi}(r,t)}.
\end{aligned}
\end{equation}

\begin{figure}
\centering
\begin{minipage}{0.42\textwidth}
\includegraphics[width=\textwidth]{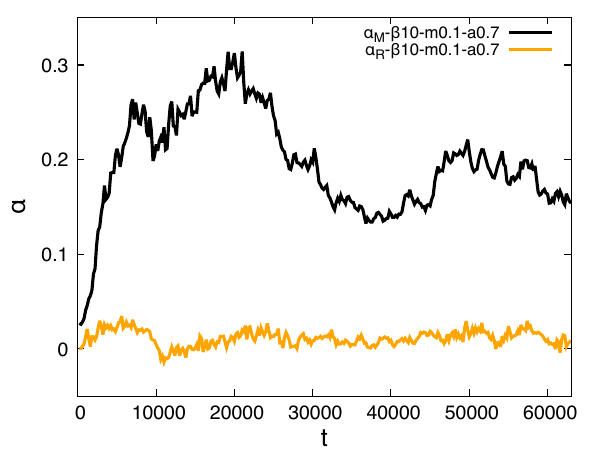}
\caption{The volume-averaged of $\alpha_M$ and $\alpha_R$ versus time for case $\beta$10-m0.1-a0.7. The $\alpha_R$ contribution to total $\alpha$ is less than 10\%.}
\label{fig:alpahM-alphaR}
\end{minipage}
\quad
\begin{minipage}{0.42\textwidth}
\includegraphics[width=\textwidth]{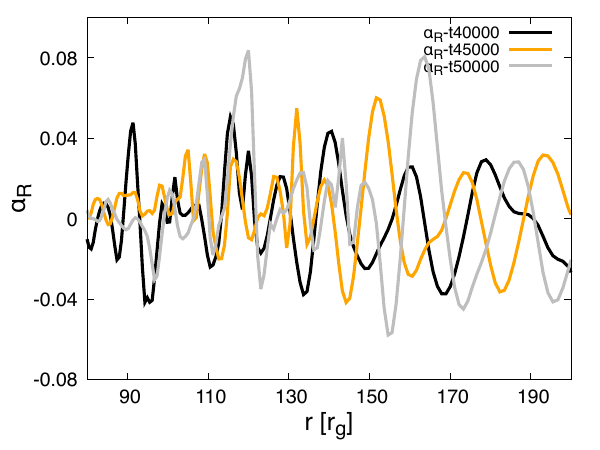}
\caption{The fluctuations of $\alpha_R$ at time snapshots $t=40000,45000,50000$ computed for $\beta$10-m0.1-a0.7. The data is zoomed-in for radii [80,200].}
\label{fig:alphaR}
\end{minipage}
\end{figure}

The average of quantity $Q$ (i.e. velocity and viscosity components) is taken vertically and weighted by density as:

\begin{equation}
    \overline{Q(r,t)} = \frac{\int_{-H}^{H} \sqrt{-g} \rho Q dz}{\int_{-H}^{H} \sqrt{-g} \rho dz}, 
    \label{eq:verticalAvg}
\end{equation}

where $H$ is the disk's scale height.
The comparison between the Maxwell and Reynolds contributions to the total volume averaged $\alpha$ over time is shown in Fig.~\ref{fig:alpahM-alphaR} for case $\beta$10-m0.1-a0.7. The $\alpha$ values are vertically averaged according to Eq.(\ref{eq:verticalAvg}). Fig.~\ref{fig:alphaR} shows the radial profiles of $\alpha_R$ at three time snapshots, $t=4 \times 10^4 ,4.5 \times 10^4 ,5 \times 10^4 $.  
These figures show that $\alpha_M$ has a bigger contribution to the averaged $\alpha$ (more than 90\%), while $\alpha_R$ fluctuates significantly due to the turbulence. At some radii, $\alpha_R$ may vary in the range of [-0.08,0.08].
The torque's fluctuation is discussed in Sec.~\ref{sec:discussion} in detail. 

\begin{figure}
\centering
\begin{minipage}{0.42\textwidth}
\includegraphics[width=\textwidth]{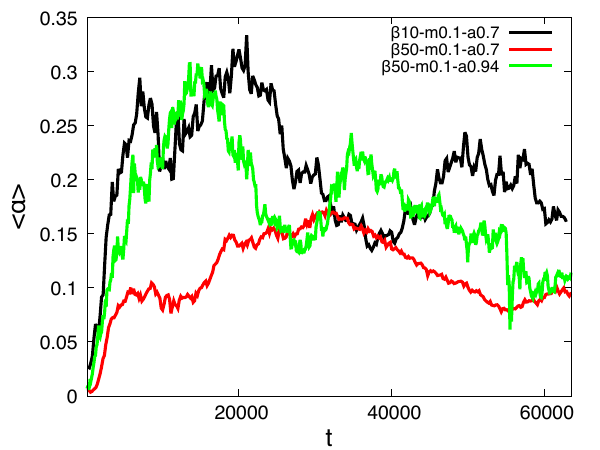}
\caption{
The volume-averaged of $\alpha$ weighted by density versus
time for $\beta$10-m0.1-a0.7, $\beta$50-m0.1-a0.7 and $\beta$50-m0.1-a0.94 cases computed inside the turbulent disk $r<150$ and $ -H < z < H$.}
\label{fig:alpha-vs-time}
\end{minipage}
\quad
\begin{minipage}{0.42\textwidth}
\includegraphics[width=\textwidth]{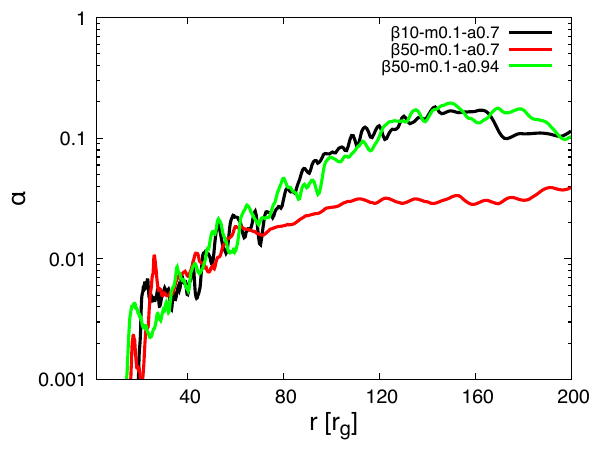}
\caption{The time averaged radial profile of total $\alpha$ for $\beta$10-m0.1-a0.7, $\beta$50-m0.1-a0.7 and $\beta$50-m0.1-a0.94 cases. The time average is taken from the second half of the evolution.}
\label{fig:radial-alpha}
\end{minipage}
\end{figure}

In Fig.~\ref{fig:alpha-vs-time} we show the volume averaged of $\alpha$ versus time for all the models except $\beta$1-m0.5-a0.7, which turns into MAD state periodically at the early time of the evolution. The volume average is taken over the turbulent part of the disk (inner radius of the grid to $r=150\,r_g$).
The comparison between the models for time averaged $\alpha$ is illustrated in Fig.~\ref{fig:radial-alpha}. The time average is taken over almost the second half of the evolution ($25000 < t < 60000$) to make sure that the MRI is being triggered and making the disk turbulent up to $r \sim 150 r_g$.
These results show that the average value of $\alpha$ may vary by factor of 2 for highly magnetized cases such as $\beta$10-m0.1-a0.7 and $\beta$50-m0.1-a0.94. It reaches 0.3 at the highest and finally converges to $\alpha \approx 0.12-0.15$ for all the cases including $\beta$50-m0.1-a0.7 at the late evolution. The average $\alpha$ drops (with a delay) in all these three cases after entering the MAD status. 
On the other hand, the radial profile in Fig.~\ref{fig:radial-alpha} demonstrates that $\alpha$ changes significantly over radius, i.e. it becomes very small at the inner radii where the MRI is suppressed, and gradually increases at the larger radii, reaching to $\alpha \approx 0.2$ at $r \sim 150 r_g$ for $\beta$50-m0.1-a0.94 and $\beta$10-m0.1-a0.7 cases, while the less magnetized case $\beta$50-m0.1-a0.7 settles down to $\alpha \approx 0.03$ for $100\,r_g < r < 200\,r_g$.  

\subsection{\label{sec:3D} Three-dimensional evolution and resolution effects}

The MRI growth rate, saturation and unstable modes are highly dependent on the grid resolution and accurate evolution of the magnetic field in 3D~\citep{Shi:2016,Oishi:2020}. 
Based on the Cowling anti-dynamo theorem, the axisymmetric magnetic fields cannot be maintained by axisymmetric motions of a conducting fluid by dynamo action~\citep{James:1980,Nunez:1996}.
On the other hand, it is well known that the disk's evolution under the MAD condition in 3D is different from 2D \citep{James:2022,White:2019,Tchekhovskoy:2011}.
To investigate the resolution and the third dimension effects we perform a high-resolution 2D simulation with 2016*1056 grid points (labelled as 2D-High), and a 3D simulation with a moderate resolution of 288*256*96 (labelled as 3D) and compare them with the standard 2D resolution (labelled as 2D-Std) of the $\beta$50-m0.1-a0.94 case. 

\begin{figure}
\centering
\includegraphics[width={0.50\textwidth}]{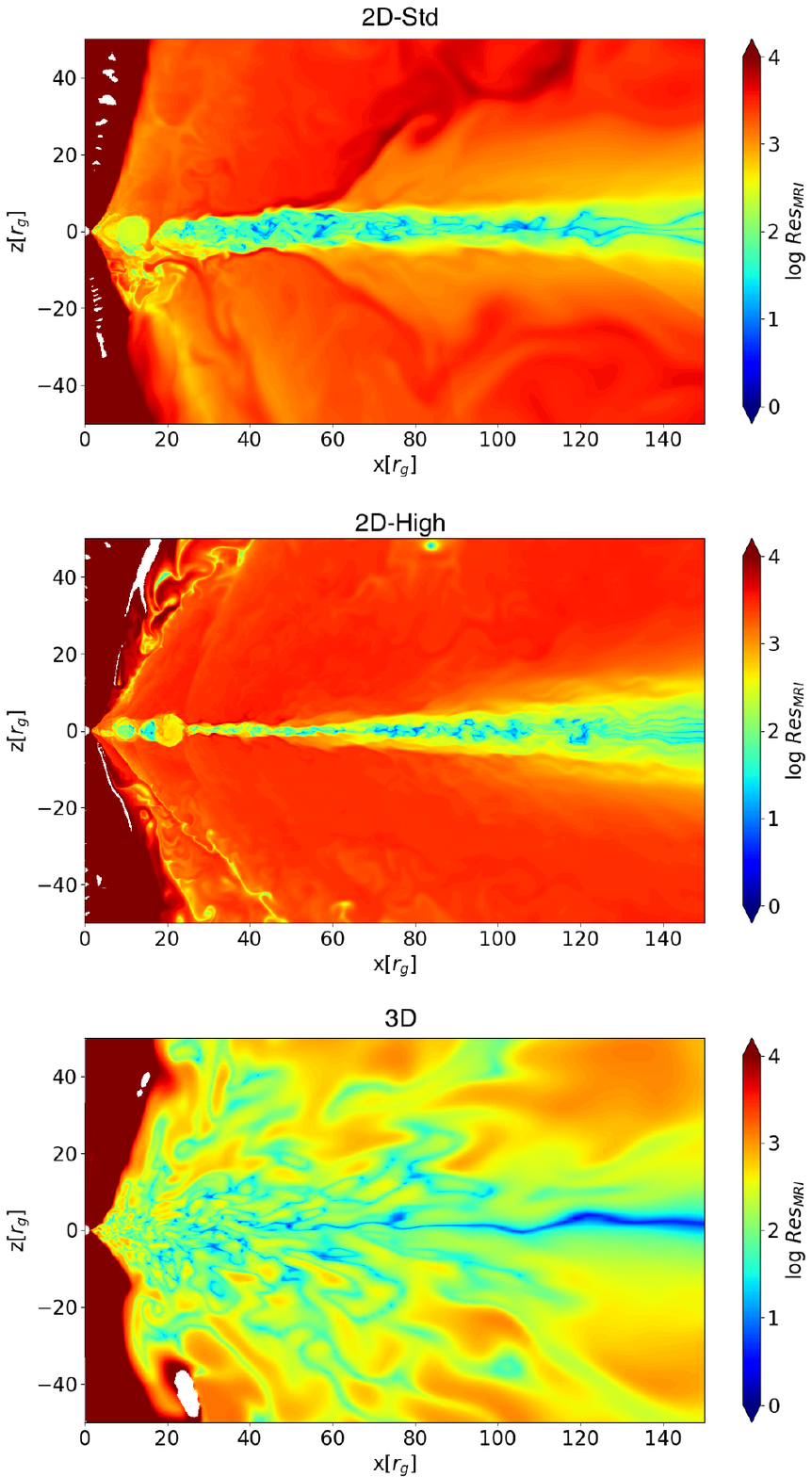}
\caption{The MRI resolution defined as $Res_{MRI}\equiv \lambda_{MRI}/\Delta \theta$ in logarithmic scale measured at $t=50000$, for 2D-Std (the standard 2D resolution grid), 2D-High (the high-resolution 2D grid) and 3D (3D grid) models with the same initial setup as $\beta$50-m0.1-a0.94. 
Almost the entire simulation domain has roughly 10 or more grid points per MRI's fastest growing wavelength, indicating MRI turbulence should be well captured.}
\label{fig:Qmri}
\end{figure}

The first numerical sanity check is whether the MRI turbulence is well captured in our simulations with the proper scaling of the cell sizes. For this purpose we compute the MRI resolution in the polar direction as ${\rm Res}_{MRI} \equiv \lambda_{MRI,\theta}/\Delta \theta$. The meridional slice of this quantity at $t=50000$ is shown in Fig.\ref{fig:Qmri} for the standard resolution of $\beta$50-m0.1-a0.94 case and compared with 3D and 2D-High test cases. We find that our grid always provides at least 10 cells per MRI fastest growing mode's wavelength in the equatorial region, where the thin disk exists, providing evidence that the MRI turbulence is well resolved within our standard 2D and 3D resolutions.

\begin{figure*}
    \centering
    \includegraphics[width=0.95\textwidth]{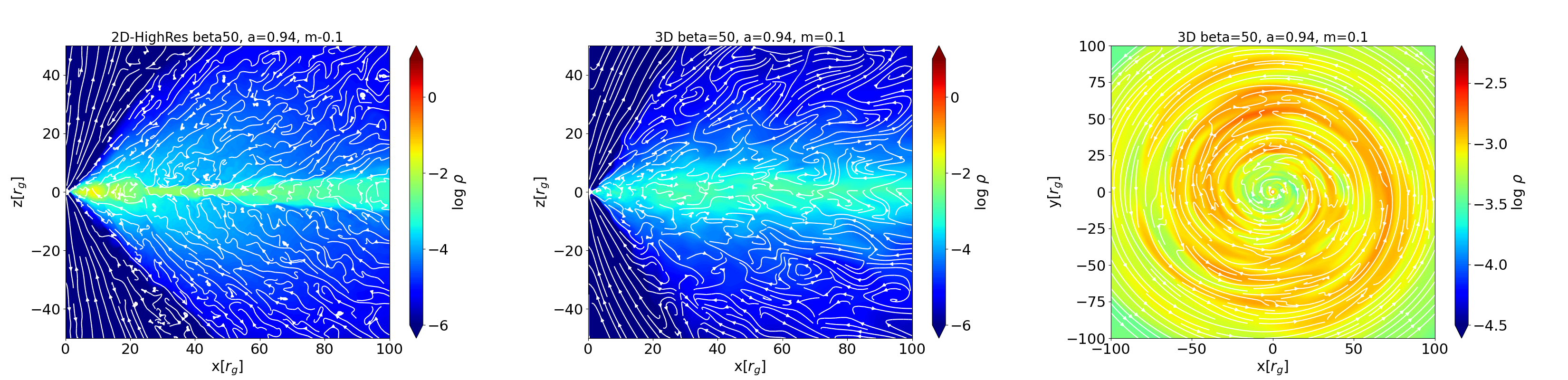}
    \caption{The density profile and magnetic field lines for 2D-High case [left], the meridional slice at $\phi=0$ [middle], and the equatorial slice of the same quantities for 3D simulation [right] at the final snapshot.}
    \label{fig:3d-slice}
\end{figure*}

\begin{figure}
\centering
\includegraphics[width={0.42\textwidth}]{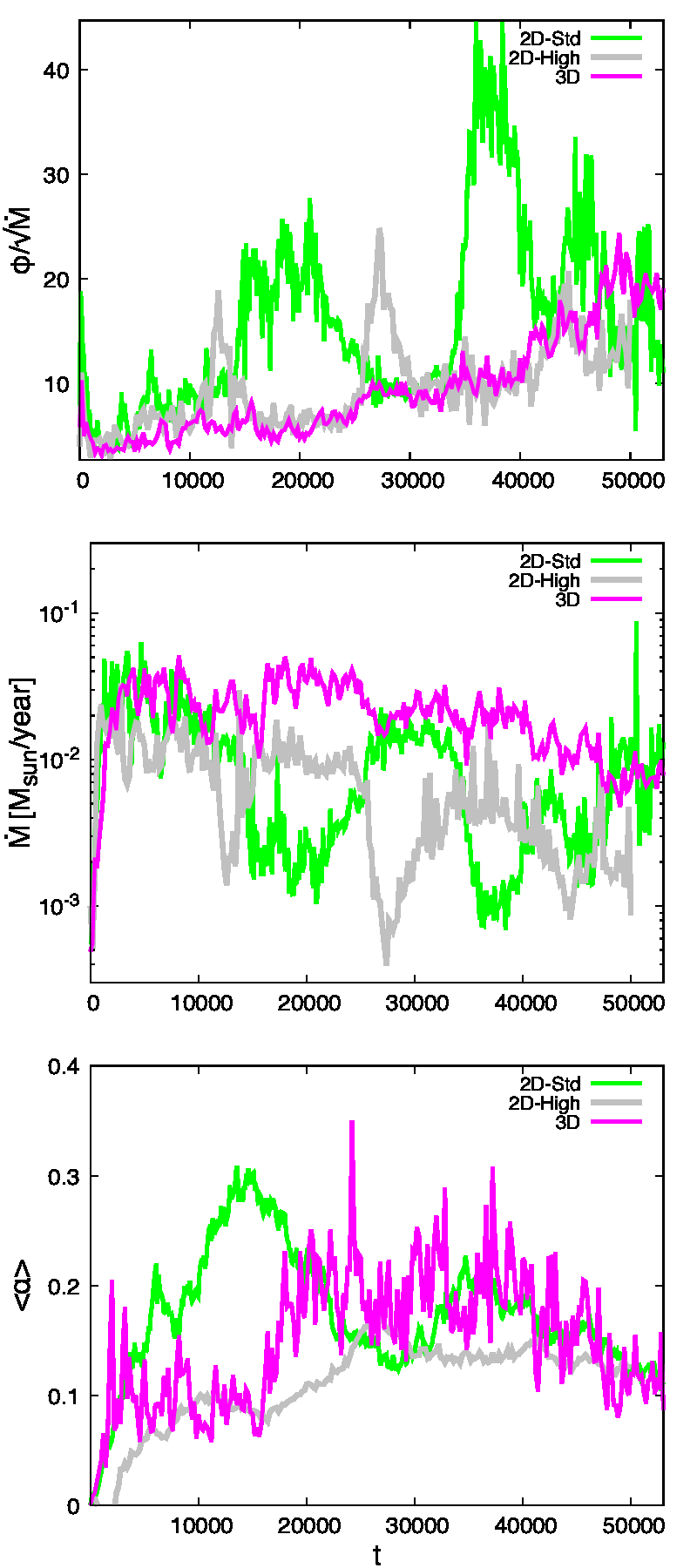}
\caption{The ratio of the magnetic flux to the square root of the mass flux on the horizon [top], the accretion rate [middle], and the volume averaged $\alpha$-viscosity measured for 2D-Std (the standard 2D resolution grid), 2D-High (the high-resolution 2D grid) and 3D (3D grid) models with the same initial setup as $\beta$50-m0.1-a0.94.}
\label{fig:3D-flux}
\end{figure}

Fig.~\ref{fig:3d-slice} shows the density and field line configuration for the 2D-High case, and the meridional and equatorial slices of the same quantities from the 3D simulation at the final time (the color scale is different in the figures). Although, in the 3D case, the disk remains geometrically thin, and the inner torus with a looping magnetic field is not formed. The other significant difference between the 2D and the 3D simulation is the accretion rate, which stays close to $0.5-1 \dot{M}_{Edd}$ for the entire simulation (see Fig.~\ref{fig:3D-flux}). It is also obvious that there is no sudden transition to the MAD state at the early evolution compared to the 2D simulation. The MAD state develops later where the $\Phi_{B}/\sqrt{\dot{M}}$ ratio gradually increases (Fig.~\ref{fig:3D-flux}), though the simulation is not long enough to observe the magnetic barrier close to the horizon and therefore causes no significant drop in the accretion rate. The 2D-High case, on the other hand, resulted in a similar qualitative evolution as the standard resolution, i. e. the highly-magnetized, plasmoid structure has been created at the inner radii, distinguishable from the turbulent thin disk with MRI channel solution at the outer radii. It also shows frequent transitions to the MAD state and significant drops in the accretion rate.

The volume averaged $\alpha$-viscosity measured for each case is shown in the bottom panel of Fig.~\ref{fig:3D-flux}\footnote{The average of the quantities in Eq.~\ref{eq:verticalAvg} are taken vertically and azimuthally for the 3D case.}. Both the 3D and high-resolution 2D cases demonstrate deviations from the standard 2D case at the earlier time of the evolution, which is expected for MRI. However, these values start converging in the middle of the simulation. The 3D evolution of $\alpha$ follows the standard 2D case closely with some fluctuations, while the average $\alpha$ for 2D-High goes through minor changes over time and it settles down to $\alpha \approx 0.12$ during the second half of the evolution. These tests show that despite of the different early evolution, our $<\alpha>$ measurements for 2D simulations are reliable for the second half of the evolution and not affected by the anti-dynamo theorem. 

\section{\label{sec:torque} Analytical results: Viscous torque measurements}

In this section, we apply the numerical results from our GR MHD simulations, reported in Sec.~\ref{sec:results}, to estimate the viscous torque in a 1D general relativistic approach. Since the secondary BH is not explicitly included in our simulation, we can make different assumptions about its mass and different scenarios it may go through while inspiraling in the turbulent gaseous environment. The results in this section are scaled for a primary BH mass of $10^6 M_{\odot}$ and mass ratio of $10^{-3}$ orbiting within the disk's inner turbulent part, at $r<200 r_g$.

In a realistic numerical model, where the secondary BH is included, one needs to calculate all the different components of the total torque exerted on
the secondary, i.e. gravitational torque, accretion on the secondary and viscous torques, to measure how much the secondary's orbit is influenced by the environment~\citep{Derdzinski:2020}. However, there are analytical approximations in the literature to estimate the torque. Historically, these approximations were applied to explain the planet radial migration in the proto-planetary disks: According to~\cite{Tanaka:2002} and~\cite{Lyra:2010} the linear torque can be estimated for two types of migration: 1- migration type-I for a very low mass companion with mass ratio ($q<10^{-4}$), and 2- migration type-II for a higher mass ratio such as $q \sim 10^{-3}$. In migration type-II the companion is massive enough to carve a low-density region (gap) inside the disk and therefore, experiences smaller torque caused by the viscosity~\citep{Lin:1986}. A similar one-dimensional migration type-II model was adopted by~\cite{Armitage:2002} for supermassive black holes merging in accretion disks, which was further advanced by~\cite{Shapiro:2013}. In the latter, the evolution equation of the surface density was derived in the curved spacetime, considering the gas effect on the secondary BH (viscous torque) and the secondary BH's gravitational tidal effects on the disk (tidal torque). Our assumption for the hypothetical mass ratio fits the migration type-II, therefore our torque estimations are limited to the viscous torque computed relativistically using Shapiro 1D approach.

In Sec.~\ref{sec:1D-GR} we present the viscous torque measurements from the time-averaged values, and in Sec.\ref{sec:fluctuation} we discuss the viscous torque fluctuations and their effects on the orbital evolution. 

\subsection{\label{sec:1D-GR} Viscous torque: 1D GR-hybrid thin disk model}

The Newtonian one-dimensional thin disk prescription used by~\cite{Garg:2022} and~\cite{Derdzinski:2020} to compute the viscous torque for migration type-II followed:

\begin{equation}
    T_{\nu,Newt} = -3\pi r^2 \Omega_2 \nu \Sigma,
    \label{eq:T-vis-Newt}
\end{equation}

where $\Omega_2 \approx \sqrt{M_p/r^3}$ is the orbital frequency of the secondary BH assuming circular orbit for the extreme mass ratio case, $M_p$ is the mass of the primary BH, $\Sigma = \int_{-H}^{H} \rho\,dz$ is the disk surface density and $\nu = \alpha c_s H$ is the kinematic viscosity.

In a relativistic formalism, we follow the simple 1D GR-hybrid model given by~\cite{Shapiro:2013}. In this approach the rest mass accretion rate due to viscous torque is

\begin{equation}
    \dot{M}_{GR} = 2\pi \left[\frac{\mathcal{G}}{\mathcal{Q}} 3 r^{1/2} \frac{\partial}{\partial r}\left(r^{1/2} \nu \Sigma_{GR} \frac{\mathcal{D}^2}{\mathcal{C}}\right) \right],
    \label{eq:mdot-GR}
\end{equation}

where $\mathcal{G}$, $\mathcal{Q}$, $\mathcal{C}$, $\mathcal{D}$ and relativistic surface density $\Sigma_{GR}$ are defined as:

\begin{equation}
\begin{aligned}
    \mathcal{G} = \mathcal{B} \mathcal{C}^{-1/2}, \\
    L^+ = M_p x \mathcal{C} (1-2ax^{-3} + a^2 x^{-4}), \\
    \mathcal{Q} = 2x^{1/2} \partial L^+ / \partial r, \\
    \mathcal{B} = 1+ax^{-3}, \\
    \mathcal{C} = 1 - 3x^{-2} + 2ax^{-3}, \\
    \mathcal{D} = 1 - 2x^{-2} + a^2x^{-4}, \\
    \Sigma_{GR} = \int_{-H}^{H} \rho u^t \sqrt{-g}\,dz, \\ 
\end{aligned}
\end{equation}

for the Kerr metric in the Boyer-Linquist coordinates, where $M_p$ and $a$ are the mass and spin of the primary BH and $x=\sqrt{r}$. The relativistic viscous torque will be computed as:

\begin{equation}
    T_{\nu,GR} = -\dot{M}_{GR} r^2 \Omega_2
    \label{eq:T-vis-GR}
\end{equation}

We get nearly identical results as from Eq.~(\ref{eq:T-vis-Newt}) in the weak field approximation, where $\mathcal{G}$, $\mathcal{Q}$, $\mathcal{C}$, and $\mathcal{D}$ approach to unity. Our results show that the Newtonian approach (Eq.\ref{eq:T-vis-Newt}) overestimates the viscous torque, especially at the inner regions, for instance, the relativistic viscous torque is $\sim$30 \% lower at $r \sim 100 r_g$. 

On the other hand, the binary is inspiraling due to losing energy by emitting gravitational waves. So, one can define the effective GW torque to be compared with the viscous torque:

\begin{equation}
    T_{GW} = \frac{1}{2} q M_p r \dot{r}_{GW} \Omega_2,
\end{equation}

where $q$ is the mass ratio and $\dot{r}_{GW}$ is derived by the quadruple approximation for the evolution of the orbital separation as follows ~\citep{Peters:1964}:

\begin{equation}
    \dot{r}_{GW} = -\frac{64}{5} \frac{(GM)^3}{c^5} \frac{1}{1+q^{-1}} \frac{1}{1+q} \frac{1}{r^3}.
\end{equation}

\begin{figure}
\centering
\includegraphics[width={0.42\textwidth}]{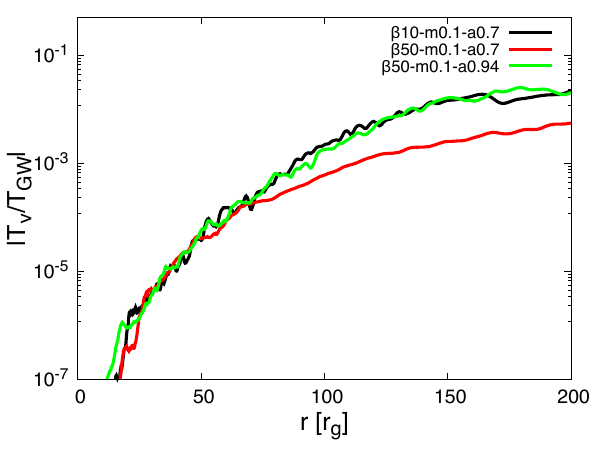}
\caption{The ratio of the viscous torque to the effective GW torque for all models scaled for fixed primary BH mass $M_p=10^6 M_{\odot}$ and mass ratio of $q=0.001$.}
\label{fig:torque-avg}
\end{figure}

Fig.~\ref{fig:torque-avg} shows the ratio $T_{\nu,GR}/T_{GW}$ computed with $q=0.001$ for different models in our simulations. The values for $\Sigma$, $\alpha$ and $c_s$ used in Eq.(\ref{eq:mdot-GR}) are taken from our simulations' time-averaged numerical measurements. This result shows that the average viscous torque may reach up to a few per cent of the gravitational torque and produce a measurable phase shift in the GW signal (for the selected mass ratio and $M_p$). 
Moreover, considering the results from $\beta$50-m0.1-a0.7 and $\beta$1-m0.5-a0.7 cases, we can claim that there are particular magnetic field strengths and configurations, which result in less effective MRI turbulence, and hence, smaller viscous torques.

In comparison with results from an analytical (and Newtonian) study done by~\cite{Garg:2022}, which considered a constant value $\alpha = 0.01$, we conclude that our numerical results provide higher values for $\alpha$. Therefore, the secondary BH experiences larger viscous torque at binary separations around $r \sim 100 r_g$. Obviously, $T_{\nu}$ is scaled by other quantities such as the surface density, sound speed and scale height. Hence, the GW phase shift can still be used to probe the density and temperature of AGN disks for $\alpha \sim 0.1$ at this binary separation. 

\subsection{\label{sec:fluctuation}Torque fluctuations}

What we estimate as a torque in the previous section is the time-averaged value of viscous torque or equivalently we can call it linear torque. However, in a realistic scenario, we do not have a one-dimensional laminar gas flow. Instead, we have to deal with nonlinear turbulent fluid which continuously interacts with the secondary orbiting BH. These nonlinear hydrodynamic interactions lead to rapid changes in density, velocity and magnetic fields, which enhance or suppress the torque value over time. The deviation from the linear torque can affect the orbital evolution of the binary and introduce an additional phase shift in GWs.~\cite{Zwick:2022} presented a detailed study on the stochastic torque or time-variable torque estimations and their measurable effects in the GW signals. As they suggested there are two important sources for these fluctuations: one is the disk-driven fluctuations experienced by the low-mass orbiting object from the turbulent fluid, and the second, is the perturber-driven fluctuations occurring due to asymmetries in the gas flow near a sufficiently massive secondary BH. 

The perturber-driven fluctuations may depend on many physical processes including the stochastic accretion and tidal effects of the secondary BH, as well as gas friction and small-scale gas dynamics. On the other hand, for fluctuation studies, one needs to measure the torque directly from a relatively long simulation in the presence of the perturber and output it frequently for the Fourier analysis. The Newtonian torque measurements and its Fourier analysis have been done by~\cite{Nelson:2005} for turbulent, magnetized protoplanetary disks. They found the torque fluctuations contain high-amplitude, low-frequency components, and the stochastic migration dominates over type-I migration for their models evolved for 100-150 planet orbits.
For a relativistic approach, one can follow the direct torque computations from~\cite{Farris:2011} (See Appendix B from~\cite{Farris:2011} for the details of the relativistic derivation of torque components).

\begin{figure}
\centering
\includegraphics[width={0.42\textwidth}]{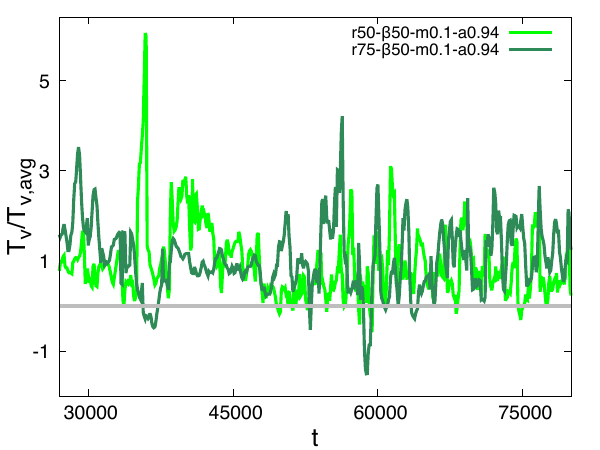}
\caption{The ratio of the viscous torque to the time-averaged viscous torque versus time at radii $r=50 r_g$ and $r=75 r_g$ for the case $\beta$50-m0.1-a0.94. The fluctuations in the measured torque affect the orbital evolution of the binary system.}
\label{fig:torque-point}
\end{figure}

In our study, the torque fluctuations do not appear in the average values we presented in Sec.~\ref{sec:1D-GR}, and overall we observed the average viscous torque is negative (by its definition) and therefore, it facilitates the shrinkage of the binary orbit. 
However, it is still interesting to investigate the fluctuations of the viscous torque and have a qualitative discussion on this topic. 
(Similar to Sec.~\ref{sec:1D-GR}, we assume that the dominant torque component is the viscous part for our chosen mass ratio, and we limit our study of the fluctuations to the viscous torque.)
Taking a close look at the case $\beta$50-m0.1-a0.94, for instance, shows that the viscous torque may deviate from the time-averaged torque by up to factor of five. Fig.~\ref{fig:torque-point} shows the ratio of the viscous torque over the time-averaged torque versus time for two selected radii ($r=50 r_g$ and $r=75 r_g$). As we observe this ratio can change dramatically over time and it even may have a different sign during evolution. Measuring a positive viscous torque may seem incorrect. However, we should remind the readers that when $\alpha$ is computed from the Reynolds and Maxwell contributions, it may take negative values by the definition in a turbulent fluid (see $\alpha_R$ in Fig.\ref{fig:alphaR} for instance). We postpone the direct computations of the torque and its fluctuations to future studies where the orbiting perturber is included in our simulations.

\section{\label{sec:discussion} Discussion: Phase shift due to gas, $\alpha$-viscosity and other important physical scenarios}

Using $\alpha$ and the linear torque computed in Sec.~\ref{sec:alpha} and~\ref{sec:torque}, we can estimate the orbital evolution and GW dephasing due to the gas environment. Here we follow the analysis given by~\cite{Zwick:2022} derived based on~\cite{Tanaka:2002,Ward:1997}, so the flux of angular momentum induced by the viscous torque is estimated from local $\alpha$, density and sound speed as:

\begin{equation}
\dot{L}_T = -\alpha \frac{6 \pi r^{7/2} c_s(r)^3 \rho(r)}{\sqrt{GM}},
\end{equation}

and we can derive the variation in the binary separation from

\begin{equation}
    \dot{r} = \dot{r}_{GW} + 2 \frac{\dot{L}_T}{Mq} \sqrt{\frac{r}{GM}} \equiv \dot{r}_{GW} + \dot{r}_T.
\end{equation}

Finally, the phase shift is approximated by

\begin{equation}
    \delta \phi = \phi_{vac} - \phi \approx 2\pi \int f_{GW}(r)\frac{\dot{r}_{gas}}{\dot{r}_{GW}^2} dr. 
\end{equation}

Approximating the GW frequency for a binary with mass ratio $0.001$ and $M_p = 10^6 M_{\odot}$ from~\cite{Derdzinski:2020}, we predict the dephasing would be roughly around $\sim 10$ radians for about $10^5$ inspiral orbits.
In comparison with the analytical work by~\cite{Garg:2022} and the numerical study by~\cite{Derdzinski:2020}, which used constant $\alpha \sim 0.01$, our computed viscous torque is slightly larger because our directly-measured $\alpha$ is larger by more than one order of magnitude. 
Generally, for LISA, there are sources with a few up to a few hundred SNR and the phase of the GW signal can be reconstructed within the accuracy of 1/SNR~\citep{Thorpe:2019}. Therefore, the predicted phase shift must be detectable by LISA for a few years of observational time~\citep{Kocsis:2011,Garg:2022,Derdzinski:2020}.

At this point, it is worth having a brief discussion over the $\alpha$-viscosity we computed and comparing it with similar works from the literature.  
Our numerical measurement for average $\alpha$-viscosity as $<\alpha>=<\alpha_{M}>+<\alpha_{R}>$ from the mid-evolution time estimates this value around $\sim 0.1-0.25$ for different cases, which varies from almost zero close to the horizon and reach up to $\sim 0.2$ at radii around $150~r_g$ after taking the time average. However, in the 1D analytical approach and viscous hydrodynamic simulations, it is very common to assume smaller constant values of the order of $0.001-0.02$~\citep{Shibata:2017} for $\alpha$. This assumption is based on high-resolution shearing box simulations presented in the literature such as~\cite{Shi:2016} and~\cite{Salvesen:2016}.
In 2007~\cite{King:2007} estimated the $\alpha$-viscosity as $\alpha \approx 0.1-0.4$ for different astrophysical systems based on the observation.
This study explained a factor of 10 discrepancy between observational
and theoretical estimates of $\alpha$ as a result of the restrictions in the local models, and suggested undertaking fully 3D, global simulations for a realistic measurement.
Since then, several convergence test studies have been done for global accretion disks including~\cite{Hawley:2013} and~\cite{Suzuki:2014}. The non-relativistic Maxwell stress ($-<B_r B_{\phi}>/4 \pi P_{gas}$) reported in~\cite{Hawley:2013} varies from 0.17 to 0.46 for different models, and what they defined as the Shakura-Sunyaev viscosity or $\alpha_{SS}$ is the density-weighted of this parameter and reported as $\alpha_{SS} \approx 0.03-0.04$ for their highest resolution. They also claimed that the resolution also affects the value of Maxwell stress or $\alpha_{mag}$, though not dramatically. Our 2D models also show slight changes in volume-integrated $\alpha$ for a different resolution, however, the relativistic definition of $\alpha$ is different in our measurements, and it includes the Reynolds stress as well. \cite{Suzuki:2014} has reported the density-weighted $\alpha_{mag}$, with the same non-relativistic definition, within the same order of magnitude ($\sim 0.01$) for 3D global disk simulation with initial vertical field configuration. 
Moreover, the most recent studies by~\cite{Mishra:2020} and~\cite{Mishra:2022} reported the value of $\alpha_{mag}$ in the range of $\sim 0.2-0.4$ for a magnetized thin disk model with $H/R \approx 0.05$. 

As our final discussion, we should emphasise that all the results presented in this work are order-of-magnitude estimations for torque measurements and their effects on the GW detections. This study is limited in many ways, most importantly for a realistic approach, one needs to include the radiative cooling for optically thin AGN disks. In such a model, radiative cooling competes with viscous heating to reach thermal equilibrium which overall affects the thermal evolution and the geometry of the disk significantly~\citep{Narayan:1998}. However, our study can still be considered as a good approach for AGNs with lower radiation efficiency.
So far, several GR MHD groups have included radiative transfer processes to study the emission spectrum from the accretion disks in their simulations including~\cite{Noble:2011,McKinney:2014,Dexter:2016,Younsi:2016,Bronzwaer:2018,Bronzwaer:2020,Moscibrodzka:2020,Chatterjee:2020,Dihingia:2023}. 
The low-mass orbiting object is another missing part of our simulation. based on the discussion given by~\cite{Zwick:2022} the perturber-driven torque fluctuations can be affected by the disk parameters and become noticeably important in GW detections.
Moreover, the presence of the low-mass secondary black hole provides an opportunity to study Lindblad or orbital resonance torques in numerical simulations.~\cite{Armitage:2002} have observed the formation of spiral waves that mediate angular momentum exchange between the secondary and the disk. It is worth mentioning that besides gas-induced torques such as the MRI-driven viscosity and Lindblad resonance, the tidally distorted primary BH's horizon can gravitationally couple to the orbiting low-mass secondary, transferring energy and angular momentum from the black hole to the orbit and cause an additional phase shift in the GW signal \citep{O'Sullivan:2014}. 

Evolving a supermassive binary black hole in magnetized fluid for hundreds of orbits (or longer) in a high-resolution grid and with all the relevant physics included (such as radiation) requires extremely expensive GR MHD simulations in dynamic spacetime. However, for future studies, it is possible to simplify the problem in the case of extreme mass ratios by evolving only MHD equations in a fixed spacetime and adding the secondary BH as a perturber. Such approximations are taken by~\cite{Combi:2021} and~\cite{Sukova:2021A} for adjusting metric and hydrodynamic equations respectively. The secondary's accretion can also be added as an additional sink term in the source part of the hydro equations (see eq.(6) from~\cite{Derdzinski:2020} as an example).

\section{\label{sec:conclusion} Summary and Conclusions}

This study was one step toward a realistic estimation of the disk's environmental effects on possible future gravitational wave detections.
We evolved several magnetized thin disk models to quantify the viscous $\alpha$ parameter in turbulent fluid developed by magnetorotational instability. We used the results of these simulations to estimate the viscous torque experienced by the hypothetical low-mass secondary BH inspiraling the primary black hole inside an AGN disk. Finally, we estimated the phase shift in the GW signal caused by the disk's environmental effect based on the torque magnitude. 

We observed the disks with well-resolved MRI have an average $\alpha$ viscosity that varies around $0.1-0.25$ during the second part of the evolution. 
The MRI is suppressed at the inner part of the disk, close to the primary BH, so the value of the $\alpha$ viscosity is negligible in this region. However, time-averaged $\alpha$ reaches $\approx 0.1-0.2$ at larger radii where the fluid is turbulent and the MRI fastest growing mode is resolvable.

Altering the initial conditions, we found that the initial magnetic field with a lower inclination angle plays an important role in triggering and sustaining MRI, while the field configuration with a higher inclination angle turns the disk into the MAD state with episodic high magnetic flux at the horizon.
The initially weakly magnetized case makes the MRI saturated at a later time and overall has smaller $\alpha$, and therefore, smaller viscous torque compared to the other cases.
Moreover, we found that the BH spin does not change the results significantly.

Since three-dimensional evolution is essential to capture all MRI's features, we carry out one case of 3D simulation. The results of this simulation show different dynamical evolution of the disk especially on developing the MAD state. Nevertheless, the average value of $\alpha$-viscosity agrees well with the 2D simulation for the long enough evolution.

We applied the numerical results from the GR MHD simulations to estimate the viscous torque using the GR-Hybrid approach for the general relativistic  one-dimensional thin disk. We found that the time-averaged viscous torque can be as large as $\sim 1\%$ of the GW torque for a mass ratio of $q=10^{-3}$ at radii around $r \sim 100 ~r_g$, where $\alpha$ is maximal. This extra torque from the environment appears as faster shrinkage of the binary's orbit and phase shift in the GW signal.
We also observed the Newtonian-calculated torque deviates from the relativist one up to 30\% higher at radii around $r \sim 100 ~r_g$.

Monitoring the viscous torque at different radii shows that the fluctuations in the torque values may change dramatically, and even sometimes it changes the torque's sign or deviates from the average value by a factor of five. The study of torque fluctuations is essential for the binary's orbital evolution and should be considered in future numerical studies where the perturbation from secondary BH is explicitly included.  

\begin{acknowledgements}
The authors thank Andrea Derdzinski, Bożena Czerny, Hector Olivares, Scott Noble, Milton Ruiz and Jose Font for helpful discussions and advice throughout this project. This work was supported by grant No. 2019/35/B/ST9/04000 from the Polish National Science Center, Poland. 
We gratefully acknowledge Polish high-performance computing infrastructure PLGrid (HPC Centers: ACK Cyfronet AGH) for providing computer facilities and support within computational grant no. PLG/2023/016071.
We also acknowledge support from the Interdisciplinary Center for Mathematical Modeling of the Warsaw University.
\end{acknowledgements}

%
   \bibliographystyle{aa} 
   \bibliography{biblio.bib} 

\begin{thebibliography}{94}
\expandafter\ifx\csname natexlab\endcsname\relax\def\natexlab#1{#1}\fi

\bibitem[{{Amaro-Seoane} {et~al.}(2023){Amaro-Seoane}, {Andrews}, {Arca Sedda},
  {Askar}, {Baghi}, {Balasov}, {Bartos}, {Bavera}, {Bellovary}, {Berry},
  {Berti}, {Bianchi}, {Blecha}, {Blondin}, {Bogdanovi{\'c}}, {Boissier},
  {Bonetti}, {Bonoli}, {Bortolas}, {Breivik}, {Capelo}, {Caramete},
  {Cattorini}, {Charisi}, {Chaty}, {Chen}, {Chru{\'s}li{\'n}ska}, {Chua},
  {Church}, {Colpi}, {D'Orazio}, {Danielski}, {Davies}, {Dayal}, {De Rosa},
  {Derdzinski}, {Destounis}, {Dotti}, {Dutan}, {Dvorkin}, {Fabj}, {Foglizzo},
  {Ford}, {Fouvry}, {Franchini}, {Fragos}, {Fryer}, {Gaspari}, {Gerosa},
  {Graziani}, {Groot}, {Habouzit}, {Haggard}, {Haiman}, {Han}, {Istrate},
  {Johansson}, {Khan}, {Kimpson}, {Kokkotas}, {Kong}, {Korol}, {Kremer},
  {Kupfer}, {Lamberts}, {Larson}, {Lau}, {Liu}, {Lloyd-Ronning}, {Lodato},
  {Lupi}, {Ma}, {Maccarone}, {Mandel}, {Mangiagli}, {Mapelli}, {Mathis},
  {Mayer}, {McGee}, {McKernan}, {Miller}, {Mota}, {Mumpower}, {Nasim},
  {Nelemans}, {Noble}, {Pacucci}, {Panessa}, {Paschalidis}, {Pfister},
  {Porquet}, {Quenby}, {Ricarte}, {R{\"o}pke}, {Regan}, {Rosswog}, {Ruiter},
  {Ruiz}, {Runnoe}, {Schneider}, {Schnittman}, {Secunda}, {Sesana}, {Seto},
  {Shao}, {Shapiro}, {Sopuerta}, {Stone}, {Suvorov}, {Tamanini}, {Tamfal},
  {Tauris}, {Temmink}, {Tomsick}, {Toonen}, {Torres-Orjuela}, {Toscani},
  {Tsokaros}, {Unal}, {V{\'a}zquez-Aceves}, {Valiante}, {van Putten}, {van
  Roestel}, {Vignali}, {Volonteri}, {Wu}, {Younsi}, {Yu}, {Zane}, {Zwick},
  {Antonini}, {Baibhav}, {Barausse}, {Bonilla Rivera}, {Branchesi},
  {Branduardi-Raymont}, {Burdge}, {Chakraborty}, {Cuadra}, {Dage}, {Davis}, {de
  Mink}, {Decarli}, {Doneva}, {Escoffier}, {Gandhi}, {Haardt}, {Lousto},
  {Nissanke}, {Nordhaus}, {O'Shaughnessy}, {Portegies Zwart}, {Pound},
  {Schussler}, {Sergijenko}, {Spallicci}, {Vernieri}, \&
  {Vigna-G{\'o}mez}}]{LISA:2023}
{Amaro-Seoane}, P., {Andrews}, J., {Arca Sedda}, M., {et~al.} 2023, Living
  Reviews in Relativity, 26, 2

\bibitem[{{Armitage} \& {Natarajan}(2002)}]{Armitage:2002}
{Armitage}, P.~J. \& {Natarajan}, P. 2002, ApJl, 567, L9

\bibitem[{{Avara} {et~al.}(2023){Avara}, {Krolik}, {Campanelli}, {Noble},
  {Bowen}, \& {Ryu}}]{Avara:2023}
{Avara}, M.~J., {Krolik}, J.~H., {Campanelli}, M., {et~al.} 2023, arXiv
  e-prints, arXiv:2305.18538

\bibitem[{{Balbus} \& {Hawley}(1991)}]{BalbusHaw1991}
{Balbus}, S.~A. \& {Hawley}, J.~F. 1991, ApJ, 376, 214

\bibitem[{{Barausse} {et~al.}(2014){Barausse}, {Cardoso}, \&
  {Pani}}]{Barausse:2014}
{Barausse}, E., {Cardoso}, V., \& {Pani}, P. 2014, \prd, 89, 104059

\bibitem[{{Barausse} \& {Rezzolla}(2008)}]{Barausse:2008}
{Barausse}, E. \& {Rezzolla}, L. 2008, \prd, 77, 104027

\bibitem[{{Bentz} {et~al.}(2007){Bentz}, {Denney}, {Cackett}, {Dietrich},
  {Fogel}, {Ghosh}, {Horne}, {Kuehn}, {Minezaki}, {Onken}, {Peterson}, {Pogge},
  {Pronik}, {Richstone}, {Sergeev}, {Vestergaard}, {Walker}, \&
  {Yoshii}}]{Bentz:2007}
{Bentz}, M.~C., {Denney}, K.~D., {Cackett}, E.~M., {et~al.} 2007, ApJ, 662, 205

\bibitem[{{Bogdanovi{\'c}} {et~al.}(2022){Bogdanovi{\'c}}, {Miller}, \&
  {Blecha}}]{Bogdanovic:2022}
{Bogdanovi{\'c}}, T., {Miller}, M.~C., \& {Blecha}, L. 2022, Living Reviews in
  Relativity, 25, 3

\bibitem[{{Bon} {et~al.}(2016){Bon}, {Zucker}, {Netzer}, {Marziani}, {Bon},
  {Jovanovi{\'c}}, {Shapovalova}, {Komossa}, {Gaskell}, {Popovi{\'c}},
  {Britzen}, {Chavushyan}, {Burenkov}, {Sergeev}, {La Mura}, {Vald{\'e}s}, \&
  {Stalevski}}]{bon}
{Bon}, E., {Zucker}, S., {Netzer}, H., {et~al.} 2016, \apjs, 225, 29

\bibitem[{{Bowen} {et~al.}(2018){Bowen}, {Mewes}, {Campanelli}, {Noble},
  {Krolik}, \& {Zilh{\~a}o}}]{Bowen:2018}
{Bowen}, D.~B., {Mewes}, V., {Campanelli}, M., {et~al.} 2018, ApJl, 853, L17

\bibitem[{{Bronzwaer} {et~al.}(2018){Bronzwaer}, {Davelaar}, {Younsi},
  {Mo{\'s}cibrodzka}, {Falcke}, {Kramer}, \& {Rezzolla}}]{Bronzwaer:2018}
{Bronzwaer}, T., {Davelaar}, J., {Younsi}, Z., {et~al.} 2018, AAp, 613, A2

\bibitem[{{Bronzwaer} {et~al.}(2020){Bronzwaer}, {Younsi}, {Davelaar}, \&
  {Falcke}}]{Bronzwaer:2020}
{Bronzwaer}, T., {Younsi}, Z., {Davelaar}, J., \& {Falcke}, H. 2020, AAp, 641,
  A126

\bibitem[{{Cardoso} {et~al.}(2022){Cardoso}, {Destounis}, {Duque}, {Macedo}, \&
  {Maselli}}]{Cardoso:2022}
{Cardoso}, V., {Destounis}, K., {Duque}, F., {Macedo}, R.~P., \& {Maselli}, A.
  2022, \prl, 129, 241103

\bibitem[{{Cardoso} \& {Maselli}(2020)}]{Cardoso:2020}
{Cardoso}, V. \& {Maselli}, A. 2020, \aap, 644, A147

\bibitem[{{Chatterjee} {et~al.}(2020){Chatterjee}, {Younsi}, {Liska},
  {Tchekhovskoy}, {Markoff}, {Yoon}, {van Eijnatten}, {Hesp}, {Ingram}, \& {van
  der Klis}}]{Chatterjee:2020}
{Chatterjee}, K., {Younsi}, Z., {Liska}, M., {et~al.} 2020, MNRAS, 499, 362

\bibitem[{{Combi} {et~al.}(2021){Combi}, {Armengol}, {Campanelli}, {Ireland},
  {Noble}, {Nakano}, \& {Bowen}}]{Combi:2021}
{Combi}, L., {Armengol}, F. G.~L., {Campanelli}, M., {et~al.} 2021, \prd, 104,
  044041

\bibitem[{{Combi} {et~al.}(2022){Combi}, {Lopez Armengol}, {Campanelli},
  {Noble}, {Avara}, {Krolik}, \& {Bowen}}]{Combi:2022}
{Combi}, L., {Lopez Armengol}, F.~G., {Campanelli}, M., {et~al.} 2022, ApJ,
  928, 187

\bibitem[{{Crenshaw} {et~al.}(2009){Crenshaw}, {Kraemer}, {Schmitt}, {Kaastra},
  {Arav}, {Gabel}, \& {Korista}}]{Crenshaw:2009}
{Crenshaw}, D.~M., {Kraemer}, S.~B., {Schmitt}, H.~R., {et~al.} 2009, ApJ, 698,
  281

\bibitem[{D'Ascoli {et~al.}(2018)D'Ascoli, Noble, Bowen, Campanelli, Krolik, \&
  Mewes}]{DAscoli:2018}
D'Ascoli, S., Noble, S.~C., Bowen, D.~B., {et~al.} 2018, Astrophys. J., 865,
  140

\bibitem[{{De Villiers} {et~al.}(2003){De Villiers}, {Hawley}, \&
  {Krolik}}]{DeVilliers:2003}
{De Villiers}, J.-P., {Hawley}, J.~F., \& {Krolik}, J.~H. 2003, ApJ, 599, 1238

\bibitem[{Derdzinski {et~al.}(2021)Derdzinski, D'Orazio, Duffell, Haiman, \&
  MacFadyen}]{Derdzinski:2020}
Derdzinski, A., D'Orazio, D., Duffell, P., Haiman, Z., \& MacFadyen, A. 2021,
  Mon. Not. Roy. Astron. Soc., 501, 3540

\bibitem[{{Dexter}(2016)}]{Dexter:2016}
{Dexter}, J. 2016, MNRAS, 462, 115

\bibitem[{{Dibi} {et~al.}(2012){Dibi}, {Drappeau}, {Fragile}, {Markoff}, \&
  {Dexter}}]{Dibi:2012}
{Dibi}, S., {Drappeau}, S., {Fragile}, P.~C., {Markoff}, S., \& {Dexter}, J.
  2012, MNRAS, 426, 1928

\bibitem[{{Dihingia} {et~al.}(2023){Dihingia}, {Mizuno}, {Fromm}, \&
  {Younsi}}]{Dihingia:2023}
{Dihingia}, I.~K., {Mizuno}, Y., {Fromm}, C.~M., \& {Younsi}, Z. 2023, arXiv
  e-prints, arXiv:2305.09698

\bibitem[{{Dihingia} {et~al.}(2021){Dihingia}, {Vaidya}, \&
  {Fendt}}]{Dihingia:2021}
{Dihingia}, I.~K., {Vaidya}, B., \& {Fendt}, C. 2021, MNRAS, 505, 3596

\bibitem[{{Duffell} {et~al.}(2014){Duffell}, {Haiman}, {MacFadyen}, {D'Orazio},
  \& {Farris}}]{Duffell:2014}
{Duffell}, P.~C., {Haiman}, Z., {MacFadyen}, A.~I., {D'Orazio}, D.~J., \&
  {Farris}, B.~D. 2014, ApJl, 792, L10

\bibitem[{{Event Horizon Telescope Collaboration} {et~al.}(2021){Event Horizon
  Telescope Collaboration}, {Akiyama}, {Algaba}, {Alberdi}, {Alef}, {Anantua},
  {Asada}, {Azulay}, {Baczko}, {Ball}, {Balokovi{\'c}}, {Barrett}, {Benson},
  {Bintley}, {Blackburn}, {Blundell}, {Boland}, {Bouman}, {Bower}, {Boyce},
  {Bremer}, {Brinkerink}, {Brissenden}, {Britzen}, {Broderick}, {Broguiere},
  {Bronzwaer}, {Byun}, {Carlstrom}, {Chael}, {Chan}, {Chatterjee},
  {Chatterjee}, {Chen}, {Chen}, {Chesler}, {Cho}, {Christian}, {Conway},
  {Cordes}, {Crawford}, {Crew}, {Cruz-Osorio}, {Cui}, {Davelaar}, {De
  Laurentis}, {Deane}, {Dempsey}, {Desvignes}, {Dexter}, {Doeleman}, {Eatough},
  {Falcke}, {Farah}, {Fish}, {Fomalont}, {Ford}, {Fraga-Encinas}, {Friberg},
  {Fromm}, {Fuentes}, {Galison}, {Gammie}, {Garc{\'\i}a}, {Gelles}, {Gentaz},
  {Georgiev}, {Goddi}, {Gold}, {G{\'o}mez}, {G{\'o}mez-Ruiz}, {Gu}, {Gurwell},
  {Hada}, {Haggard}, {Hecht}, {Hesper}, {Himwich}, {Ho}, {Ho}, {Honma},
  {Huang}, {Huang}, {Hughes}, {Ikeda}, {Inoue}, {Issaoun}, {James}, {Jannuzi},
  {Janssen}, {Jeter}, {Jiang}, {Jimenez-Rosales}, {Johnson}, {Jorstad}, {Jung},
  {Karami}, {Karuppusamy}, {Kawashima}, {Keating}, {Kettenis}, {Kim}, {Kim},
  {Kim}, {Kim}, {Kino}, {Koay}, {Kofuji}, {Koch}, {Koyama}, {Kramer}, {Kramer},
  {Krichbaum}, {Kuo}, {Lauer}, {Lee}, {Levis}, {Li}, {Li}, {Lindqvist}, {Lico},
  {Lindahl}, {Liu}, {Liu}, {Liuzzo}, {Lo}, {Lobanov}, {Loinard}, {Lonsdale},
  {Lu}, {MacDonald}, {Mao}, {Marchili}, {Markoff}, {Marrone}, {Marscher},
  {Mart{\'\i}-Vidal}, {Matsushita}, {Matthews}, {Medeiros}, {Menten}, {Mizuno},
  {Mizuno}, {Moran}, {Moriyama}, {Moscibrodzka}, {M{\"u}ller}, {Musoke}, {Mus
  Mej{\'\i}as}, {Michalik}, {Nadolski}, {Nagai}, {Nagar}, {Nakamura},
  {Narayan}, {Narayanan}, {Natarajan}, {Nathanail}, {Neilsen}, {Neri}, {Ni},
  {Noutsos}, {Nowak}, {Okino}, {Olivares}, {Ortiz-Le{\'o}n}, {Oyama},
  {{\"O}zel}, {Palumbo}, {Park}, {Patel}, {Pen}, {Pesce}, {Pi{\'e}tu},
  {Plambeck}, {PopStefanija}, {Porth}, {P{\"o}tzl}, {Prather},
  {Preciado-L{\'o}pez}, {Psaltis}, {Pu}, {Ramakrishnan}, {Rao}, {Rawlings},
  {Raymond}, {Rezzolla}, {Ricarte}, {Ripperda}, {Roelofs}, {Rogers}, {Ros},
  {Rose}, {Roshanineshat}, {Rottmann}, {Roy}, {Ruszczyk}, {Rygl},
  {S{\'a}nchez}, {S{\'a}nchez-Arguelles}, {Sasada}, {Savolainen}, {Schloerb},
  {Schuster}, {Shao}, {Shen}, {Small}, {Sohn}, {SooHoo}, {Sun}, {Tazaki},
  {Tetarenko}, {Tiede}, {Tilanus}, {Titus}, {Toma}, {Torne}, {Trent},
  {Traianou}, {Trippe}, {van Bemmel}, {van Langevelde}, {van Rossum}, {Wagner},
  {Ward-Thompson}, {Wardle}, {Weintroub}, {Wex}, {Wharton}, {Wielgus}, {Wong},
  {Wu}, {Yoon}, {Young}, {Young}, {Younsi}, {Yuan}, {Yuan}, {Zensus}, {Zhao},
  \& {Zhao}}]{EHT2021}
{Event Horizon Telescope Collaboration}, {Akiyama}, K., {Algaba}, J.~C.,
  {et~al.} 2021, ApJl, 910, L13

\bibitem[{Farris {et~al.}(2011)Farris, Liu, \& Shapiro}]{Farris:2011}
Farris, B.~D., Liu, Y.~T., \& Shapiro, S.~L. 2011, Phys. Rev. D, 84, 024024

\bibitem[{{Franchini} {et~al.}(2023{\natexlab{a}}){Franchini}, {Bonetti},
  {Lupi}, {Miniutti}, {Bortolas}, {Giustini}, {Dotti}, {Sesana}, {Arcodia}, \&
  {Ryu}}]{Franchini:2023QPE}
{Franchini}, A., {Bonetti}, M., {Lupi}, A., {et~al.} 2023{\natexlab{a}}, \aap,
  675, A100

\bibitem[{{Franchini} {et~al.}(2022){Franchini}, {Lupi}, \&
  {Sesana}}]{Franchini:2022}
{Franchini}, A., {Lupi}, A., \& {Sesana}, A. 2022, ApJl, 929, L13

\bibitem[{{Franchini} {et~al.}(2023{\natexlab{b}}){Franchini}, {Lupi},
  {Sesana}, \& {Haiman}}]{Franchini:2023}
{Franchini}, A., {Lupi}, A., {Sesana}, A., \& {Haiman}, Z. 2023{\natexlab{b}},
  MNRAS, 522, 1569

\bibitem[{Frank {et~al.}(2002)Frank, King, \& Raine}]{frank_king_raine_2002}
Frank, J., King, A., \& Raine, D. 2002, ACCRETION DISCS, 3rd edn. (Cambridge
  University Press), 80–151

\bibitem[{{Gammie} {et~al.}(2003){Gammie}, {McKinney}, \&
  {T{\'o}th}}]{Gammie2003}
{Gammie}, C.~F., {McKinney}, J.~C., \& {T{\'o}th}, G. 2003, ApJ, 589, 444

\bibitem[{{Garg} {et~al.}(2022){Garg}, {Derdzinski}, {Zwick}, {Capelo}, \&
  {Mayer}}]{Garg:2022}
{Garg}, M., {Derdzinski}, A., {Zwick}, L., {Capelo}, P.~R., \& {Mayer}, L.
  2022, MNRAS, 517, 1339

\bibitem[{{Graham} {et~al.}(2015){Graham}, {Djorgovski}, {Stern}, {Glikman},
  {Drake}, {Mahabal}, {Donalek}, {Larson}, \& {Christensen}}]{Graham:2015}
{Graham}, M.~J., {Djorgovski}, S.~G., {Stern}, D., {et~al.} 2015, \nat, 518, 74

\bibitem[{{Hawley} {et~al.}(2013){Hawley}, {Richers}, {Guan}, \&
  {Krolik}}]{Hawley:2013}
{Hawley}, J.~F., {Richers}, S.~A., {Guan}, X., \& {Krolik}, J.~H. 2013, \apj,
  772, 102

\bibitem[{{Igumenshchev}(2008)}]{Igumenshchev:2008}
{Igumenshchev}, I.~V. 2008, ApJ, 677, 317

\bibitem[{{Ingram} {et~al.}(2021){Ingram}, {Motta}, {Aigrain}, \&
  {Karastergiou}}]{ingram21}
{Ingram}, A., {Motta}, S.~E., {Aigrain}, S., \& {Karastergiou}, A. 2021, MNRAS,
  503, 1703

\bibitem[{{James} {et~al.}(2022){James}, {Janiuk}, \& {Nouri}}]{James:2022}
{James}, B., {Janiuk}, A., \& {Nouri}, F.~H. 2022, ApJ, 935, 176

\bibitem[{{James} {et~al.}(1980){James}, {Winch}, \& {Roberts}}]{James:1980}
{James}, R.~W., {Winch}, D.~E., \& {Roberts}, P.~H. 1980, Geophysical and
  Astrophysical Fluid Dynamics, 15, 149

\bibitem[{{Janiuk}(2017)}]{Janiuk2017}
{Janiuk}, A. 2017, ApJ, 837, 39

\bibitem[{{Janiuk}(2019)}]{Janiuk-2019}
{Janiuk}, A. 2019, ApJ, 882, 163

\bibitem[{{King} {et~al.}(2007){King}, {Pringle}, \& {Livio}}]{King:2007}
{King}, A.~R., {Pringle}, J.~E., \& {Livio}, M. 2007, \mnras, 376, 1740

\bibitem[{Kocsis {et~al.}(2011)Kocsis, Yunes, \& Loeb}]{Kocsis:2011}
Kocsis, B., Yunes, N., \& Loeb, A. 2011, Phys. Rev. D, 84, 024032

\bibitem[{{Lin} \& {Papaloizou}(1986)}]{Lin:1986}
{Lin}, D.~N.~C. \& {Papaloizou}, J. 1986, ApJ, 309, 846

\bibitem[{{Linial} \& {Metzger}(2023)}]{Linial-Metzger:2023}
{Linial}, I. \& {Metzger}, B.~D. 2023, \apj, 957, 34

\bibitem[{{Liska} {et~al.}(2020){Liska}, {Tchekhovskoy}, \&
  {Quataert}}]{Liska:2020}
{Liska}, M., {Tchekhovskoy}, A., \& {Quataert}, E. 2020, MNRAS, 494, 3656

\bibitem[{{Lyra} {et~al.}(2010){Lyra}, {Paardekooper}, \& {Mac
  Low}}]{Lyra:2010}
{Lyra}, W., {Paardekooper}, S.-J., \& {Mac Low}, M.-M. 2010, ApJl, 715, L68

\bibitem[{{Mahesh} {et~al.}(2023){Mahesh}, {McWilliams}, \&
  {Pirog}}]{Mahesh:2023}
{Mahesh}, S., {McWilliams}, S.~T., \& {Pirog}, M. 2023, arXiv e-prints,
  arXiv:2305.01533

\bibitem[{{McKinney} {et~al.}(2012){McKinney}, {Tchekhovskoy}, \&
  {Blandford}}]{McKinney:2012}
{McKinney}, J.~C., {Tchekhovskoy}, A., \& {Blandford}, R.~D. 2012, MNRAS, 423,
  3083

\bibitem[{{McKinney} {et~al.}(2014){McKinney}, {Tchekhovskoy}, {Sadowski}, \&
  {Narayan}}]{McKinney:2014}
{McKinney}, J.~C., {Tchekhovskoy}, A., {Sadowski}, A., \& {Narayan}, R. 2014,
  MNRAS, 441, 3177

\bibitem[{{Milosavljevi{\'c}} \& {Merritt}(2003)}]{Milosavljevic:2003}
{Milosavljevi{\'c}}, M. \& {Merritt}, D. 2003, in American Institute of Physics
  Conference Series, Vol. 686, The Astrophysics of Gravitational Wave Sources,
  ed. J.~M. {Centrella}, 201--210

\bibitem[{{Miniutti} {et~al.}(2023){Miniutti}, {Giustini}, {Arcodia}, {Saxton},
  {Chakraborty}, {Read}, \& {Kara}}]{Miniutti:2023}
{Miniutti}, G., {Giustini}, M., {Arcodia}, R., {et~al.} 2023, \aap, 674, L1

\bibitem[{{Miniutti} {et~al.}(2019){Miniutti}, {Saxton}, {Giustini},
  {Alexander}, {Fender}, {Heywood}, {Monageng}, {Coriat}, {Tzioumis}, {Read},
  {Knigge}, {Gandhi}, {Pretorius}, \& {Ag{\'\i}s-Gonz{\'a}lez}}]{Miniutti:2019}
{Miniutti}, G., {Saxton}, R.~D., {Giustini}, M., {et~al.} 2019, \nat, 573, 381

\bibitem[{{Mishra} {et~al.}(2020){Mishra}, {Begelman}, {Armitage}, \&
  {Simon}}]{Mishra:2020}
{Mishra}, B., {Begelman}, M.~C., {Armitage}, P.~J., \& {Simon}, J.~B. 2020,
  \mnras, 492, 1855

\bibitem[{{Mishra} {et~al.}(2022){Mishra}, {Fragile}, {Anderson},
  {Blankenship}, {Li}, \& {Nalewajko}}]{Mishra:2022}
{Mishra}, B., {Fragile}, P.~C., {Anderson}, J., {et~al.} 2022, \apj, 939, 31

\bibitem[{Moody {et~al.}(2019)Moody, Shi, \& Stone}]{Moody:2019}
Moody, M. S.~L., Shi, J.-M., \& Stone, J.~M. 2019, Astrophys. J., 875, 66

\bibitem[{{Mo{\'s}cibrodzka}(2020)}]{Moscibrodzka:2020}
{Mo{\'s}cibrodzka}, M. 2020, MNRAS, 491, 4807

\bibitem[{{Narayan} {et~al.}(2003){Narayan}, {Igumenshchev}, \&
  {Abramowicz}}]{Narayan:2003}
{Narayan}, R., {Igumenshchev}, I.~V., \& {Abramowicz}, M.~A. 2003, PASJ, 55,
  L69

\bibitem[{{Narayan} {et~al.}(1998){Narayan}, {Mahadevan}, \&
  {Quataert}}]{Narayan:1998}
{Narayan}, R., {Mahadevan}, R., \& {Quataert}, E. 1998, in Theory of Black Hole
  Accretion Disks, ed. M.~A. {Abramowicz}, G.~{Bj{\"o}rnsson}, \& J.~E.
  {Pringle}, 148--182

\bibitem[{{Narayan} \& {Yi}(1994)}]{Narayan&Yi:1994}
{Narayan}, R. \& {Yi}, I. 1994, ApJl, 428, L13

\bibitem[{{Nelson}(2005)}]{Nelson:2005}
{Nelson}, R.~P. 2005, AAp, 443, 1067

\bibitem[{{Noble} {et~al.}(2006){Noble}, {Gammie}, {McKinney}, \& {Del
  Zanna}}]{Noble2006}
{Noble}, S.~C., {Gammie}, C.~F., {McKinney}, J.~C., \& {Del Zanna}, L. 2006,
  ApJ, 641, 626

\bibitem[{{Noble} {et~al.}(2011){Noble}, {Krolik}, {Schnittman}, \&
  {Hawley}}]{Noble:2011}
{Noble}, S.~C., {Krolik}, J.~H., {Schnittman}, J.~D., \& {Hawley}, J.~F. 2011,
  ApJ, 743, 115

\bibitem[{{Noble} {et~al.}(2012){Noble}, {Mundim}, {Nakano}, {Krolik},
  {Campanelli}, {Zlochower}, \& {Yunes}}]{Noble:2012}
{Noble}, S.~C., {Mundim}, B.~C., {Nakano}, H., {et~al.} 2012, ApJ, 755, 51

\bibitem[{Nunez(1996)}]{Nunez:1996}
Nunez, M. 1996, SIAM Review, 38, 553

\bibitem[{{Oishi} {et~al.}(2020){Oishi}, {Vasil}, {Baxter}, {Swan}, {Burns},
  {Lecoanet}, \& {Brown}}]{Oishi:2020}
{Oishi}, J.~S., {Vasil}, G.~M., {Baxter}, M., {et~al.} 2020, Proceedings of the
  Royal Society of London Series A, 476, 20190622

\bibitem[{{O'Sullivan} \& {Hughes}(2014)}]{O'Sullivan:2014}
{O'Sullivan}, S. \& {Hughes}, S.~A. 2014, \prd, 90, 124039

\bibitem[{{Peters}(1964)}]{Peters:1964}
{Peters}, P.~C. 1964, Physical Review, 136, 1224

\bibitem[{{Ripperda} {et~al.}(2022){Ripperda}, {Liska}, {Chatterjee}, {Musoke},
  {Philippov}, {Markoff}, {Tchekhovskoy}, \& {Younsi}}]{Ripperda:2022}
{Ripperda}, B., {Liska}, M., {Chatterjee}, K., {et~al.} 2022, ApJl, 924, L32

\bibitem[{{Saade} {et~al.}(2023){Saade}, {Brightman}, {Stern}, {Connor},
  {Djorgovski}, {D'Orazio}, {Ford}, {Graham}, {Haiman}, {Jun}, {Kammoun},
  {Kraft}, {McKernan}, {Vikhlinin}, \& {Walton}}]{Saade:2023}
{Saade}, M.~L., {Brightman}, M., {Stern}, D., {et~al.} 2023, arXiv e-prints,
  arXiv:2304.06144

\bibitem[{{Saade} {et~al.}(2020){Saade}, {Stern}, {Brightman}, {Haiman},
  {Djorgovski}, {D'Orazio}, {Ford}, {Graham}, {Jun}, {Kraft}, {McKernan},
  {Vikhlinin}, \& {Walton}}]{Saade:2020}
{Saade}, M.~L., {Stern}, D., {Brightman}, M., {et~al.} 2020, ApJ, 900, 148

\bibitem[{{Salvesen} {et~al.}(2016){Salvesen}, {Simon}, {Armitage}, \&
  {Begelman}}]{Salvesen:2016}
{Salvesen}, G., {Simon}, J.~B., {Armitage}, P.~J., \& {Begelman}, M.~C. 2016,
  \mnras, 457, 857

\bibitem[{{Sapountzis} \& {Janiuk}(2019)}]{Sap2019}
{Sapountzis}, K. \& {Janiuk}, A. 2019, ApJ, 873, 12

\bibitem[{{Shakura} \& {Sunyaev}(1976)}]{Shakura:1976}
{Shakura}, N.~I. \& {Sunyaev}, R.~A. 1976, MNRAS, 175, 613

\bibitem[{{Shapiro}(2013)}]{Shapiro:2013}
{Shapiro}, S.~L. 2013, \prd, 87, 103009

\bibitem[{{Shapovalova} {et~al.}(2004){Shapovalova}, {Doroshenko}, {Bochkarev},
  {Burenkov}, {Carrasco}, {Chavushyan}, {Collin}, {Vald{\'e}s}, {Borisov},
  {Dumont}, {Vlasuyk}, {Chilingarian}, {Fioktistova}, \&
  {Martinez}}]{shapovalova}
{Shapovalova}, A.~I., {Doroshenko}, V.~T., {Bochkarev}, N.~G., {et~al.} 2004,
  \aap, 422, 925

\bibitem[{{Shi} {et~al.}(2016){Shi}, {Stone}, \& {Huang}}]{Shi:2016}
{Shi}, J.-M., {Stone}, J.~M., \& {Huang}, C.~X. 2016, \mnras, 456, 2273

\bibitem[{{Shibata} {et~al.}(2017){Shibata}, {Kiuchi}, \&
  {Sekiguchi}}]{Shibata:2017}
{Shibata}, M., {Kiuchi}, K., \& {Sekiguchi}, Y.-i. 2017, \prd, 95, 083005

\bibitem[{{Speri} {et~al.}(2023){Speri}, {Antonelli}, {Sberna}, {Babak},
  {Barausse}, {Gair}, \& {Katz}}]{Speri:2023}
{Speri}, L., {Antonelli}, A., {Sberna}, L., {et~al.} 2023, Physical Review X,
  13, 021035

\bibitem[{{Stone} \& {Norman}(1994)}]{Stone-Norman:1994}
{Stone}, J.~M. \& {Norman}, M.~L. 1994, ApJ, 433, 746

\bibitem[{{Sukov{\'a}} {et~al.}(2021){Sukov{\'a}}, {Zaja{\v{c}}ek}, {Witzany},
  \& {Karas}}]{Sukova:2021A}
{Sukov{\'a}}, P., {Zaja{\v{c}}ek}, M., {Witzany}, V., \& {Karas}, V. 2021, ApJ,
  917, 43

\bibitem[{{Suzuki} \& {Inutsuka}(2014)}]{Suzuki:2014}
{Suzuki}, T.~K. \& {Inutsuka}, S.-i. 2014, \apj, 784, 121

\bibitem[{{Tagawa} \& {Haiman}(2023)}]{Tagawa-Haiman:2023}
{Tagawa}, H. \& {Haiman}, Z. 2023, \mnras, 526, 69

\bibitem[{{Tanaka} {et~al.}(2002){Tanaka}, {Takeuchi}, \& {Ward}}]{Tanaka:2002}
{Tanaka}, H., {Takeuchi}, T., \& {Ward}, W.~R. 2002, ApJ, 565, 1257

\bibitem[{{Tchekhovskoy} {et~al.}(2011){Tchekhovskoy}, {Narayan}, \&
  {McKinney}}]{Tchekhovskoy:2011}
{Tchekhovskoy}, A., {Narayan}, R., \& {McKinney}, J.~C. 2011, MNRAS, 418, L79

\bibitem[{{Thorpe} {et~al.}(2019){Thorpe}, {Ziemer}, {Thorpe}, {Livas},
  {Conklin}, {Caldwell}, {Berti}, {McWilliams}, {Stebbins}, {Shoemaker},
  {Ferrara}, {Larson}, {Shoemaker}, {Key}, {Vallisneri}, {Eracleous},
  {Schnittman}, {Kamai}, {Camp}, {Mueller}, {Bellovary}, {Rioux}, {Baker},
  {Bender}, {Cutler}, {Cornish}, {Hogan}, {Manthripragada}, {Ware},
  {Natarajan}, {Numata}, {Sankar}, {Kelly}, {McKenzie}, {Slutsky}, {Spero},
  {Hewitson}, {Francis}, {DeRosa}, {Yu}, {Hornschemeier}, \&
  {Wass}}]{Thorpe:2019}
{Thorpe}, J.~I., {Ziemer}, J., {Thorpe}, I., {et~al.} 2019, in Bulletin of the
  American Astronomical Society, Vol.~51, 77

\bibitem[{{Tiede} {et~al.}(2020){Tiede}, {Zrake}, {MacFadyen}, \&
  {Haiman}}]{Tiede:2020}
{Tiede}, C., {Zrake}, J., {MacFadyen}, A., \& {Haiman}, Z. 2020, ApJ, 900, 43

\bibitem[{{Ward}(1997)}]{Ward:1997}
{Ward}, W.~R. 1997, ICARUS, 126, 261

\bibitem[{{White} {et~al.}(2019){White}, {Stone}, \& {Quataert}}]{White:2019}
{White}, C.~J., {Stone}, J.~M., \& {Quataert}, E. 2019, ApJ, 874, 168

\bibitem[{Younsi {et~al.}(2016)Younsi, Zhidenko, Rezzolla, Konoplya, \&
  Mizuno}]{Younsi:2016}
Younsi, Z., Zhidenko, A., Rezzolla, L., Konoplya, R., \& Mizuno, Y. 2016, Phys.
  Rev. D, 94, 084025

\bibitem[{Yunes {et~al.}(2011)Yunes, Kocsis, Loeb, \& Haiman}]{Yunes:2011}
Yunes, N., Kocsis, B., Loeb, A., \& Haiman, Z. 2011, Phys. Rev. Lett., 107,
  171103

\bibitem[{{Zanni} {et~al.}(2007){Zanni}, {Ferrari}, {Rosner}, {Bodo}, \&
  {Massaglia}}]{Zanni:2007}
{Zanni}, C., {Ferrari}, A., {Rosner}, R., {Bodo}, G., \& {Massaglia}, S. 2007,
  AAp, 469, 811

\bibitem[{{Zwick} {et~al.}(2022){Zwick}, {Derdzinski}, {Garg}, {Capelo}, \&
  {Mayer}}]{Zwick:2022}
{Zwick}, L., {Derdzinski}, A., {Garg}, M., {Capelo}, P.~R., \& {Mayer}, L.
  2022, MNRAS, 511, 6143

\end{thebibliography}
%

\end{document}